\documentclass[superscriptaddress,
 twocolumn,amsmath,amssymb,
 aps,
prl,
floatfix,
]{revtex4-1}
\usepackage{amssymb}
\usepackage{mathrsfs}
\usepackage{graphicx}
\usepackage{dcolumn}
\usepackage{bm}
\usepackage{amsmath}
\usepackage{amsfonts}
\usepackage{color}
\usepackage{float}
\usepackage{graphicx}
\usepackage{dcolumn}
\usepackage{bm}
\usepackage{hyperref}

\usepackage{xcolor}
\hypersetup{
	colorlinks,
	linkcolor={red!50!black},
	citecolor={green!50!black},
	urlcolor={blue!80!black}
}
\usepackage[caption=false]{subfig}
\makeatletter
\newcommand{\colorcaption}[2][]{%
  \begingroup%
  \renewcommand{\@caption@fignum@sep}{ (color online). }%
  \caption[#1]{#2}%
  \endgroup%
}
\makeatother

\begin{document}


\title{Entangled 4D multi-component topological states from photonic crystal defects}

\author{Xiao Zhang}
 \email{zhangxiao@mail.sysu.edu.cn}
\affiliation{School of Physics, Sun Yat-sen University, Guangzhou 510275, China}

\author{Youjian Chen}
\affiliation{School of Physics, Sun Yat-sen University, Guangzhou 510275, China}
\affiliation{Department of physics and astronomy, Stony Brook University ,Stony Brook,NY 11794-3800}

  \author{Yuzhu Wang}
\affiliation{School of Physics, Sun Yat-sen University, Guangzhou 510275, China}

 \author{Yuhan Liu}
\affiliation{School of Physics, Sun Yat-sen University, Guangzhou 510275, China}
\affiliation{ Department of Physics, University of Chicago, 5720 S. Ellis. Avenue, Chicago, IL 60637}

 \author{Jun Yu Lin}
\affiliation{Department of Physics, The Chinese University of Hong Kong, Hong Kong, China}

 \author{Nai Chao Hu}
 \affiliation{Department of Physics, University of Texas at Austin, Austin, TX 78712, USA}
  \author{Bochen Guan}
\affiliation{Department of Electrical and Computer Engineering, University of Wisconsin, Madison, WI 53706, USA}
\author{Ching Hua Lee}
 \email{calvin-lee@ihpc.a-star.edu.sg}
\affiliation{Institute of High Performance Computing, 138632, Singapore}
\affiliation{Department of Physics, National University of Singapore, Singapore, 117542.}

\date{\today}

\begin{abstract}
Recently, there has been a drive towards the realization of topological phases beyond conventional electronic materials, including phases defined in more than three dimensions. We propose a versatile and experimentally realistic approach of realizing a large variety of multi-component topological phases in 2D photonic crystals with quasi-periodically modulated defects. With a length scale introduced by a background resonator lattice, the defects are found to host various effective orbitals of $s$, $p$ and $d$-type symmetries, thus providing a monolithic platform for realizing multi-component topological states without requiring separate internal degrees of freedom in the physical setup. Notably, by coupling the defect modulations diagonally, we report the novel realization of ``entangled'' 4D QH phase which cannot be factorized into two copies of 2D QH phases each described by the 1st Chern number. The structure of this non-factorizability can be quantified by a classical entanglement entropy inspired by quantum information theory. In another embodiment, we present 4D p-orbital nodal lines in a nonsymmorphic photonic lattice, hosting boundary states with an exotic manifold. Our simple and versatile approach hold the promise of novel topological optoelectronic and photonic applications such as one-way optical fibers.

\end{abstract}

\maketitle


\noindent\textit{Introduction--}
Quantum Hall systems~\cite{von1986quantized,novoselov2007room,jain1989composite}, topological insulators~\cite{moore2010birth,hasan2010colloquium,qi2011topological,zhang2012actinide,MHua2016} and Weyl semimetals~\cite{wan2011prb,WeylsemimetalHongMing,WeylsemimetalHasan} rank amongst the most intensely studied topics in condensed matter physics~\cite{moore2009topological,zhang2010topological,lee2015geometric,lee2015negative}. Each energy band is ascribed a topological index that mandate the presence of interesting boundary states and quantized responses. Soon after their discovery in electronic materials, analogs in electrical~\cite{ningyuan2015time,lee2017topolectrical,wang2018topologically,imhof2018topolectrical,lu2018probing,helbig2018band,hofmann2019chiral,lee2019imaging}, phononic (acoustic)~\cite{susstrunk2015observation,nash2015topological,wang2015topological,He2016,ong2016transport,lee2017topological}, photonic~\cite{haldane2008possible,wang2009observation,lin2017line} or their hybrids~\cite{peano2015topological} have been proposed and even realized experimentally. Although topological phases with nontrivial Chern invariants in principle exist in any even number of dimensions, physical space is limited by three dimensions (3D). Topological phases characterized by the second Chern number $C_2$ (4D quantum Hall (QH) systems~\cite{zhang2001four}), as well as 4D nodal band structures, cannot be directly realized within 3D physical space~\footnote{Realizations of 4D circuits in 3D~\cite{li2019boundary} may be arguable exceptions, although there will still not exist four orthogonal physical dimensions. }, and have for a long time been believed to exist only in theory.

Recently, such 4D QH phases were realized in 2D artificial systems with additional synthetic dimensions. Proposals include uni-directional propagating waveguides~\cite{kraus2013four,Zilberberg2018}, cold atoms~\cite{price2015four,price2016measurement,lohse2018}, optical fibers of Weyl materials with helical structure~\cite{lu2016topological} and an artificially generated parameter space of a four-level system~\cite{sugawa2016observation}. In the former two, synthetic dimensions arise by identifying the aperiodic hamiltonian with a Hofstadter lattice containing twice the number of dimensions~\cite{hatsugai1993edge,dean2013hofstadter,madsen2013topological,tran2015topological,fuchs2016hofstadter}. The resultant fractal band structure can be characterized by the $C_2$ in the combined space of the physical and synthetic dimensions. Experiments have been successful in realizing such quasiperiodic topological phases, with topological pumping by the first Chern number ($C_1$) observed in propagating waveguides~\cite{kraus2012topological}, and pumping by the 2nd Chern number (4D QH analog) recently reported in photonic~\cite{Zilberberg2018} and cold atoms systems~\cite{lohse2018}.

So far, however, all theoretical and experimental quasi-periodic 4D QH systems in the literature have been restricted to 2D tensor products of two 1D Hofstadter models. As such, the 4D QH states obtained were always factorizable into a product of 2D QH states, whose properties are already well-known. Specifically, the spectra of such 4D tensor product states are just the sum of two 2D QH spectra, whose spectral flow can always be understood via iterated 1st Chern number polarization~\cite{yu2011equivalent,qi2011generic,lee2014lattice}.

As such, our goal is to propose experimentally realistic setups that host truly nontrivial 4D QH modes that cannot be understood via the 1st Chern number, i.e. \emph{entangled} 4D quantum Hall modes with unremovable classical entanglement between the two 2D sub-sectors that they exist in. By computing the entanglement spectrum of various cuts, as defined in analogy to quantum entanglement, insights can be gleaned about how the sub-sectors are coupled. A fundamental observation is that the Wannier center response of entangled 4D systems cannot be expressed in terms of separate 2D Hall responses, as commonly taken for granted in the higher dimension QH literature~\cite{price2015four,petrides20186d,lee2018electromagnetic}. Our physical setup consists of a regular 2D background photonic crystal (PC) interspersed by defect resonators possessing aperiodically modulated radii. With their mature fabrication technology~\cite{Quasicrystal2,Quasicrystal3,Quasicrystal1,Quasicrystal4,Maser1,Maser2,Maser3,villeneuve1996microcavities,PhotonicApplication1,PhotonicApplication2,PhotonicApplication3}, such PCs can be easily fabricated to possess arbitrarily modulated defects, with practical applications like one-way optical fibers. This freedom in modulation is not only crucial for realizing the (classically) entangled 4D QH modes, but also yields higher $p,d$ effective orbitals that leads to multi-component phases far beyond the conventional Hofstadter model, like 4D p-orbital nodal lines in a nonsymmorphic photonic lattice.



\begin{figure}
\centering
\includegraphics[width=\linewidth]{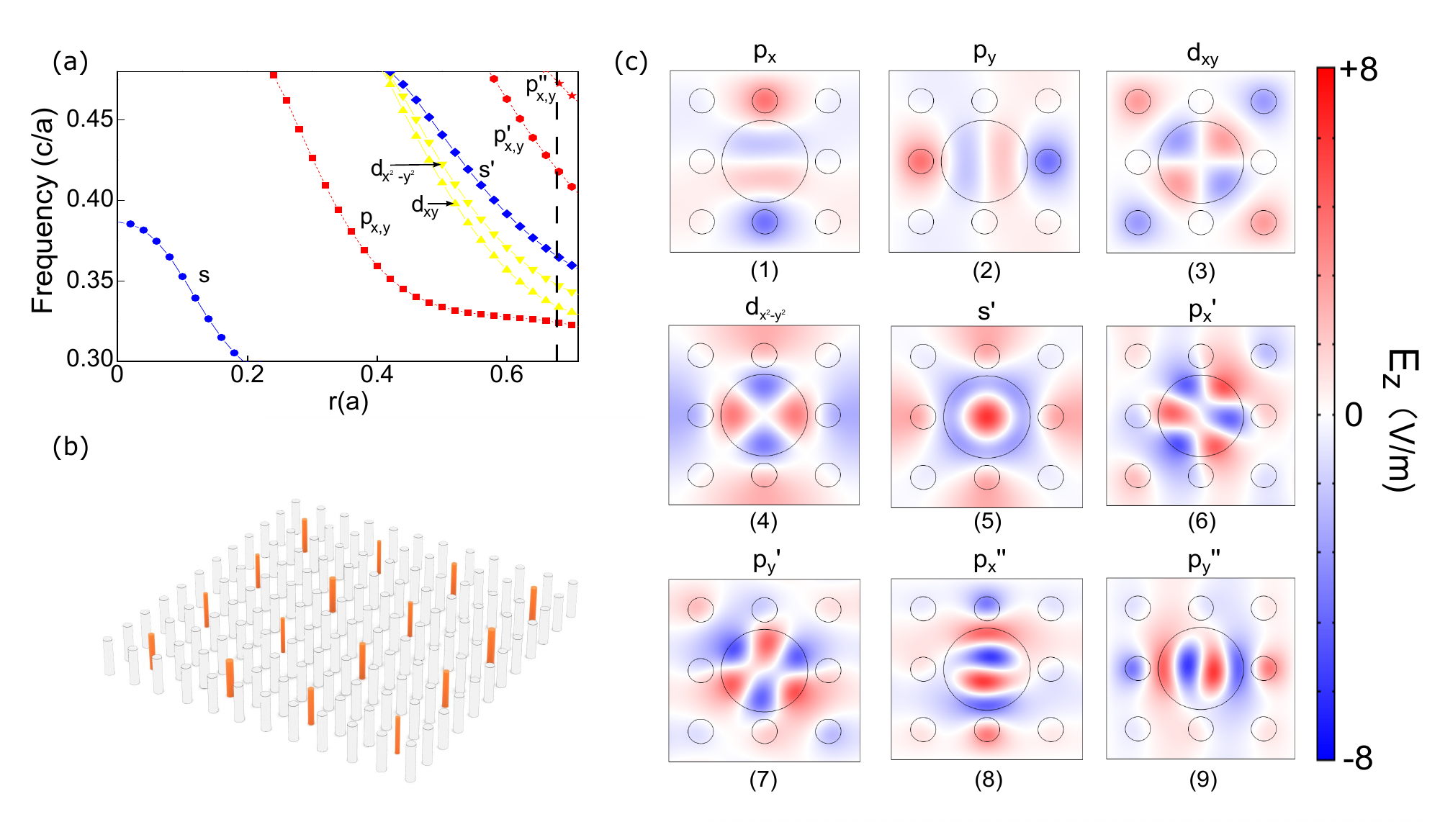}
\caption{ (a) Dependence of the onsite frequencies of various $s,p$ and $d$-type defect modes with the defect rod radius $r$. Only $s$-type modes exist for $r<0.2a$, the background rod radius, but progressively more exotic modes appear at larger radii. (b) Illustration of photonic crystal (PC) with background/defect rods colored grey/orange. Slight spatial modulation of the defect radii leads to emergent entangled 4D QH state described in Fig.~\ref{fig:4dqh}. (c) The $E_z$ distributions of the various modes in (b). The proliferation of these modes lead to interesting multi-component effective Hamiltonians. }
\label{fig:sdefect}
\end{figure}

\noindent\textit{Tight-binding Hamiltonian of a defect lattice --}
Our photonic crystal (PC) contains a background lattice of equally spaced identical dielectric rods separated by air. Importantly, embedded in the PC are ``defect'' rods of quasi-periodically modulated radii. As shown in Fig. \ref{fig:sdefect}, these defect rods act as resonators that host confined transverse magnetic (TM) electromagnetic modes, which serve as photonic analogs to the atomic orbitals of a conventional material. To identify our PC with a quasi-periodic topological lattice, we express Maxwell's equation in the basis of the TM defect modes, such that it plays the role of the Hamiltonian in the time-dependent Schr\"{o}dinger's equation~\cite{wang2005magneto,fang2011microscopic}\:
\begin{equation}
H|\psi\rangle =\begin{bmatrix}
0 & i\varepsilon_{0}^{-1}\vec\nabla\times\\
-i\mu_{0}^{-1}\vec\nabla\times & 0\\	
\end{bmatrix}\begin{bmatrix}
\vec{E}\\
\vec{H}
\end{bmatrix}=\omega|\psi\rangle
\end{equation}
where $H$ is the effective TB Hamiltonian, $|\psi\rangle =\begin{bmatrix}
\vec{E}\\
\vec{H}
\end{bmatrix}$
denotes an eigenmode with eigenfrequency $\omega$, and $\varepsilon_0$ is the spatially varying dielectric constant. We shall always set the background rods to have relative dielectric constant $8$ and radius $0.2a$, where the lattice spacing $a$ is defined as the distance between adjacent defect rods.

The shape of a defect mode depend qualitatively on the defect rod radius. For different radii, the defect modes can exhibit $s,p$ or $d$-type symmetries~\footnote{Similar behavior have been observed in other contexts previously~\cite{villeneuve1996microcavities} have demonstrated similar behavior.} (Fig.~\ref{fig:sdefect} and \cite{SOM}), hence providing a monolithic platform for constructing multi-component topological states without the need for multiple physical ``orbitals''. As illustrated in Fig.~\ref{fig:sdefect}, even with defect rods (central circles) of circular cross section, various defect modes with $s,p$ or $d$ symmetries can be realized at sufficiently large rod radii. For these modes to robustly serve as topological degrees of freedom, we require them to lie within the frequency gap of the background resonators, which from our simulations (see SOM~\cite{SOM}) range from $\omega\approx 0.32 c/a$ to $\omega\approx 0.46 c/a$, with $c$ being the speed of light. Within this frequency window, defect rods thinner than the background rods ($r<0.2a$) can support only $s$-type orbitals (Fig.~\ref{fig:sdefect}a). However, thicker defect rods can support not just $p_x$ or $p_y$-type orbitals, but also simultaneously more exotic $d_{x^2-y^2},d_{xy},p'_x,p'_y$ etc. orbitals (see SOM~\cite{SOM} for their field distributions) at sufficiently large radius. It is only through the length scale introduced by the background rods that we obained very localized qualitatively distinct $s,p$ and $d$ modes.  


For simplicity, we shall mostly focus on multi-component Hamiltonians spanned by only $s$,$p$-type defect modes, which respectively arise from defect rods with radii $r<0.2a$ and $r>0.2a$. Due to their locality, uncoupled $s$ and $p$-type defect modes form nearly flat bands with eigenfrequencies almost perfectly proportional to the defect radius (see SOM~\cite{SOM}). As such, they collectively describe give rise to an effective tight-binding (TB) Hamiltonian~\cite{wang2005magneto,Hamiltonian1} with on-site energies $\omega_i$ proportional to the radius of the $i$-th rod, and almost constant hopping amplitudes $t_{ij}$ between nearest neighborhood orbital sites $|\psi_i^\alpha\rangle,|\psi_j^\beta\rangle$:
\begin{equation}
H=\sum_{i}\sum_{\alpha\in s,p,d}\omega_i^\alpha|\psi_{i}^\alpha\rangle \langle \psi_{i}^\alpha|+\sum_{\langle i,j\rangle}\sum_{(\alpha,\beta)\in s,p,d }t_{ij}^{\alpha\beta}|\psi_{i}^\alpha\rangle \langle \psi_{j}^\beta|
\label{tb1}
\end{equation}
Further hoppings can be neglected as the defect modes are localized.
For $s$-type modes, it suffices to approximate $t^s_{ij}$ by a constant $t$; whereas for $p$-modes, $t^p_{ij}\approx t_{\sigma}$ or $t_{\pi}$ depending on whether $\sigma$ or $\pi$ bonding is involved.

\noindent\textit{Entangled 4D QH modes --}
Since the topological pumping of 2D QH states can be realized in the 1D lines of defect discussed above, we should be able to probe 4D QH states in 2D defect lattices (Fig.~\ref{fig:sdefect}b) with 2 synthetic dimensions. A 4D QH state is characterized by a nonzero
\begin{eqnarray}
C_2&=&\frac1{8\pi^2}\int F\wedge F ~d^4k\notag\\
&=& \frac1{4\pi^2}\int F_{zx}F_{yw}+F_{xy}F_{zw}+F_{yz}F_{xw}~d^4k
\label{C2}
\end{eqnarray}
where $F_{\mu\nu}=\langle \partial_{k_\mu}\phi|\partial_{k_\nu}\phi\rangle-\langle \partial_{k_\nu}\phi|\partial_{k_\mu}\phi\rangle$ is the (2D) Berry curvature for a state $\phi$ which also gives rise to a first Chern number $C_1^{\mu\nu}=\frac1{2\pi}\int F_{\mu\nu}\,d^2k$. Due to its quadratic dependence on $F_{\mu\nu}$, the 4D QH effect possess interesting nonlinear response properties\cite{PhysRevLett.115.195303} and gives the parent state for various descendent 2D and 3D topological states\cite{qi2008topological}.

The simplest approach to photonic analog of a 4D QH state is to consider independent defect modulations in both directions, i.e. $r_{ind}(x,y)=r_0+r_1 \cos (2\pi bx+k_z)+r_2\cos(2\pi by+k_w)$ (see \cite{SOM} for detailed results), as is performed with cold atoms in the literature~\cite{kraus2012topological,madsen2013topological}. However, a key limitation of this approach is that its resultant eigenmodes are just \emph{product states} of two 1st Chern number eigenmodes. Therefore, the topological pumping of such a product 4D QH mode can be completely understood in terms of the 1st Chern number, despite formally possessing a 2nd Chern number invariant.

\begin{figure}
\centering
\includegraphics[width=.85\linewidth]{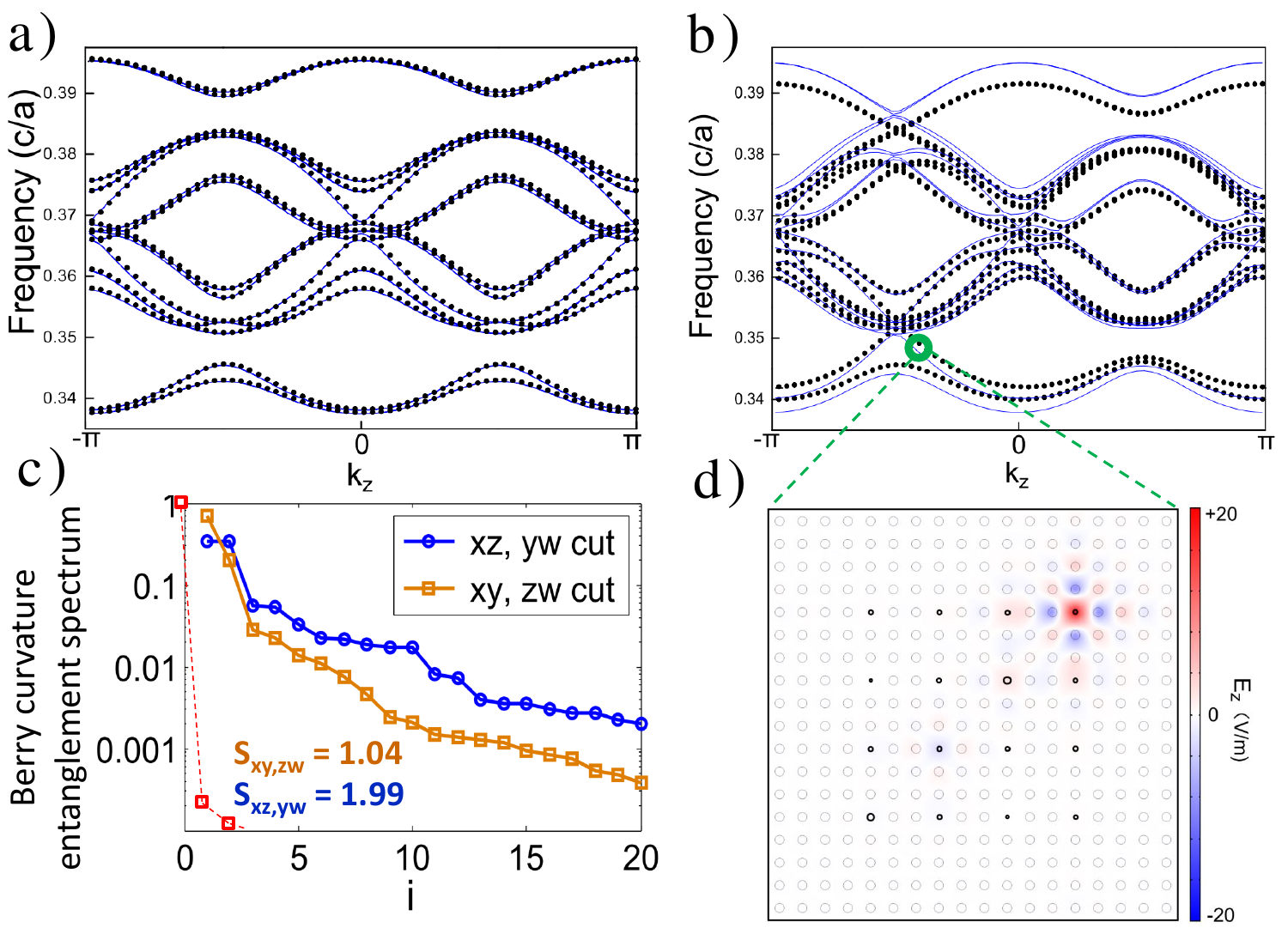}
\caption{PBC (a) and OBC (b) frequency dispersion of the entangled defect lattice of Eq.~\ref{eq4}, with parameters $r_{0}=0.1a$, $r_{1}=r_{2}=0.025a$, $k_w=0$ and rational flux $b=1/4$. Excellent numerical agreement exist between the FEM simulation/TB model (Eq.~\ref{TB2} with  with $n=4$ along $x$ and $y$. By fitting the band data with Eq.~\ref{TB2}, we found that $\omega_0=0.36862(c/a)$, $\lambda=-0.014508$ and $t=-0.00398(c/a)$) results, which are represented by the dotted/continuous curves. With OBC, topological spectral flow exists between the central and top/bottom defect bulk bands. (c) The Berry curvature entanglement spectra $\tilde\lambda_i$ and entropy for both $xy,zw$ and $xz,yw$ cuts, with long tails indicative of nontrivial entanglement, i.e. non-factorizability of the 2nd Chern number. An un-entangled defect lattice given by $r_{ind}$ gives a delta-function like spectrum (red). (d) The $E_z$ distribution of the mid-gap mode circled green ($k_z=-1.2$,$\omega=0.3465(c/a)$) in (b), which is clearly localized at a corner defect rod (thick circle). }
\label{fig:4dqh}
\end{figure}

To obtain a truly nontrivial 4D QH state analog, the key insight is to instead consider \emph{"entangled"} modulations, the simplest of which takes the form $r(x,y)=$
\begin{equation}
r_{0}+r_{1} \cos(2\pi b_z (x+y)+k_z) +r_{2} \cos(2\pi b_w (x-y)+k_w)
\label{eq4}
\end{equation}
where $r_0,r_1,r_2$ are constants, and $x,y$ labels the defect sites in the $\hat x,\hat y$-direction (Fig.~\ref{fig:4dqh}d). Analogous to the 1D AAH model~\cite{SOM}, $b_z$ and $b_w$ are flux parameters conjugate to synthetic momenta $k_z$ and $k_w$. Notably, this system is not factorizable into a direct product of two AAH models, despite the seemingly naive replacement $(x,y)\rightarrow (x+y,x-y)$.  To concretely see why, we examine its effective TB Hamiltonian
\begin{equation}
\begin{split}
H_{\text{4D}}&=\sum_{x,y}(\omega_0+\lambda \cos(2\pi b_z(x+y)+k_z)\left|\psi_{(x,y)}\right\rangle \left\langle \psi_{(x,y)}\right|\\
&+\lambda\sum_{x,y} \cos(2\pi b_w(x-y)+k_w)\left|\psi_{(x,y)}\right\rangle \left\langle \psi_{(x,y)}\right|\\
&+t\sum_{x,y} \left( \left|\psi_{(x,y)}\right\rangle \left\langle \psi_{(x+ 1,y)}\right|+\left|\psi_{(x,y)}\right\rangle \left\langle \psi_{(x,y+ 1)}\right|+h.c.\right)
\end{split}
\label{TB2}
\end{equation}
possessing $t$-hoppings in the $\hat x$ and $\hat y$ directions which are \emph{not} aligned with the phase factors of $2\pi b(x\,\pm \,y)$ along $k_z$ and $k_w$ directions. In other words, while the effective Landau gauge is taken with respect to the $\frac{\pi}{4}$-rotated directions $\hat x\pm \hat y$, the physical positions of the defect rods are still arranged in a lattice spanned by $\hat x$ and $\hat y$ basis vectors. As such, Eq.~\ref{eq4} and \ref{TB2} represent a 4D QH Hamiltonian with novel nontrivial entanglement between its two AAH subsystems. The frequency spectrum of the defect modes can be computed by a finite-element Maxwell's equation (FEM) solver, and is presented, for $b=1/4$, in Fig.~\ref{fig:4dqh}a,b for $s$-type defect modes with perodic/open boundary condition (PBC/OBC) in the $\hat x,\hat y$-direction. Upon varying $k_y$ over a period, boundary modes in the OBC case (Fig.~\ref{fig:4dqh}d) are seen to continuously connect the bulk bands. This spectral flow of mid-gap states is signature of topological behavior, and can indeed be understood via the effective TB Hamiltonian for $s$-type orbitals.

This (classical) entanglement within $H_{\text{4D}}$ can be quantified in analogy to quantum entanglement. Generically, we define an entanglement cut by decomposing a given state space $\Sigma$ as a product of two desired subspaces: $\Sigma=\Sigma_1\otimes \Sigma_2$.  For a particular state $\phi$, how much the degrees of freedom (DOFs) of the two subsystems $\Sigma_1$ and $\Sigma_2$ are entangled can be expressed through the singular-value-decomposition (SVD)
\begin{equation}
\phi=\sum_i\lambda_i\,\phi^i_1\otimes \phi^i_2
\label{SVD}
\end{equation}
where $\phi_\alpha^i$, $\alpha=1,2$ is nonzero only within $\Sigma_\alpha$.  $\lambda_1\geq\lambda_2\geq \lambda_3...$ are known as the entanglement eigenvalues, and are normalized according to $\sum_i\lambda_i=1$. In the special case of $\phi$ being a product state, there will only be one nonzero $\lambda_i$, i.e. $\lambda_1=1$ and $\lambda_{i\geq 2}=0$. The amount of entanglement, or extent of departure from a product state, is measured through the \emph{entanglement entropy}
\begin{equation}
S=-\sum_i\lambda_i\log\lambda_i\geq 0
\label{S}
\end{equation}
through which the rank, or effective number of independent DOFs, is given by $e^S\geq 1$.

Whether the 2nd Chern number $C_2$ is trivially the product of two $C_1$s (as in previous literature) depends crucially on the factorizability of the integrand in Eq.~\ref{C2}. In the purely factorizable case dictated by $r_{ind}$, the independence of the $x,k_z$ DOFs from the $y,k_w$ DOFs imply that only the $F_{zx}F_{yw}$ term is nonzero, and that $C_2=\frac1{(2\pi)^2}\int F_{zx}dk_xdk_z\int F_{yw}dk_ydw_z=C_1^{zx}C_1^{yw}$. But with the two AAH subsystems entangled like in Eq.~\ref{eq4}, the integrand for $C_2$ becomes nonfactorizable, albeit SVD decomposable into a linear combination of factorizable terms. Hence we can define the \emph{Berry curvature entanglement spectrum} $\{\tilde\lambda_1,\tilde\lambda_2,...\}$ and (classical) \emph{entanglement entropy} $S_{xz,yw}$ of a 2D defect lattice ($H_{\text{4D}}$) via Eqs.~\ref{SVD} and \ref{S}, with $\phi$ being the integrand of $C_2$ and the cut demarcating the two coupled AAH subsystems with  $x,k_z$ and $y,k_w$ DOFs.

As plotted in Fig.~\ref{fig:4dqh}c, $H_{\text{4D}}$ possess a long tail of Berry curvature entanglement eigenvalues representing not one but $e^{S_{xz,yw}}\approx 7.3$ effective entangled DOFs. Also presented in Fig.~\ref{fig:4dqh}c is the entanglement $S_{xy,zw}$ between the spatial ($x,y$) and magnetic translation $(k_z,k_w)$ DOFs, which possess a double degeneracy in the largest few eigenvalues. 
Interestingly, the Berry curvature entanglement eigenvalues $\tilde\lambda_i$ can be expressed in terms of the entanglement eigenvalues\footnote{This entanglement between different sectors of the \emph{classical} defect band should not be confused with many-body entanglement between sub-sectors of a \emph{quantum} system.}  $\lambda_i$ of the \emph{defect band} as well as its SVD-decomposed Berry connection and curvature, as detailed in the SOM~\cite{SOM}.

Since the defect lattices are quasi-periodic, the Berry curvatures are most directly computed by implementing the momentum conjugate to the quasi-periodicity as a threaded flux through an edge~\cite{fukui2005chern} (see SOM~\cite{SOM}). With the defect lattice given by Eq.~\ref{eq4}, the lowest two defect bands intersect (Fig.~\ref{fig:4dqh}), giving rise to a \emph{non-abelian} 2nd Chern number $C_2=\frac1{4\pi^2}\int F_{zx}(k)F_{yw}(k)+F_{xy}(k)F_{zw}(k)+F_{yz}(k)F_{xw}(k)~d^4k= 1+0+1=2$.

This expression for $C_2$, which interprets 4D electromagnetic response as iterated Hall responses with the Berry flux and magnetic field~\cite{SOM}, suggests an important physical consequence of entanglement to Wannier center evolution, the electromagnetic pumping of a maximally localized occupied state. As elaborated in the SOM~\cite{SOM}, the Wannier response coefficient in the generic \emph{entangled} case is $\frac1{(4\pi^2)^3}\int F_{zx}(k)F_{yw}(k')+F_{xy}(k)F_{zw}(k')+F_{yz}(k)F_{xw}(k')~d^4kd^4k'$, a much more complicated expression  that does not reduce to $C_2$. The discrepancy arises from couplings between the two QH copies which introduces 4D dispersion to Berry flux.

\noindent\textit{4D nodal lines with nonsymmorphic symmetry --} The propensity for our photonic setup in realizing entangled phases extends beyond QH states. Consider the rather extreme but perfectly realistic case with rod ``defects" being their anisotropy. As detailed below, various interesting 4D nodal line phases can arise when the modulations and orientations of these elliptical rods obey nonsymmorphic (glide reflection) symmetry, which has very much been associated with exotic topological phenomena~\cite{wang2016hourglass} in materials and metamaterials~\cite{parameswaran2013topological,schoop2016dirac,lin2017line,liu2018topological}.

\begin{figure}
\centering
\includegraphics[width=\linewidth]{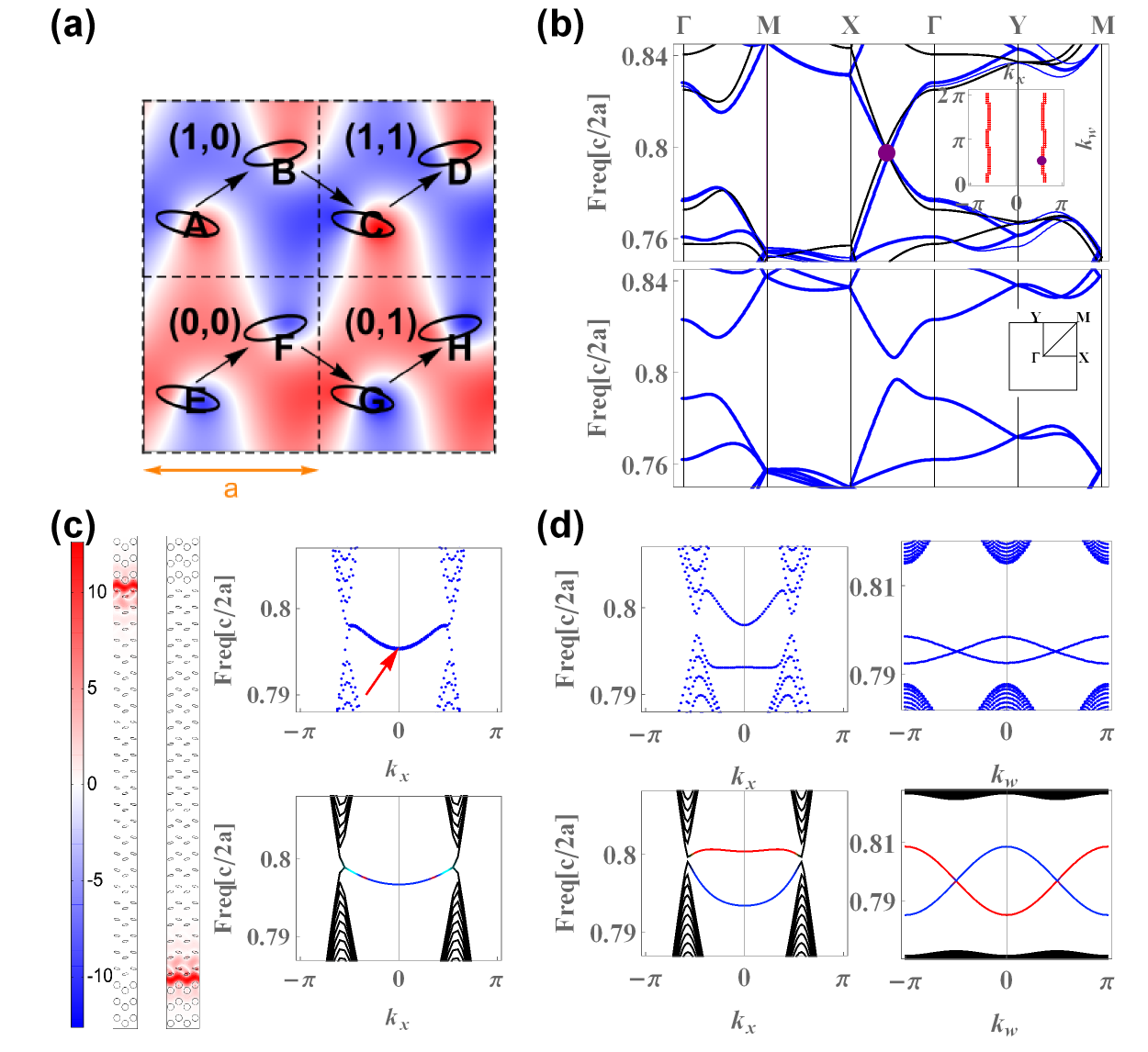}
\caption{(a) Unit cell of our effectively 4d PC with two synthetic dimensions, with red/blue representing positive/negative electric fields . Nonsymmorhic symmetry holds when $k_z=\pm\pi/2$, as indicated by arrows that map the elliptical rods to themselves via the glide reflection $g_y$. (b) Bulk band structure from COMSOL simulations (blue) and TB model (black) in the $k_z=k_w=\pi/2$ slice, with nonsymmorphic symmetry preserved/broken by tilting half of the rods (Top/Bottom). Top: A Dirac point degeneracy (purple dot) lies on nodal lines (red) in the 4D BZ, as plotted in the $k_x$-$k_w$ slice with $k_z=\pi/2$, $k_y=0$ (inset). Bottom: Degeneracies are all gapped by breaking nonsymmorphic symmetry. (c) Degenerate boundary states in the $k_z=k_w=\pi/2$ slice under y-direction OBCs.  Left: Spatial profile of boundary degeneracies at $k_x=0$, marked by the red arrow in the COMSOL band structure (Right Top), which corroborates with TB results (Right Bottom). (d) Boundary modes that do not merge into the bulk, shown for slices $k_z=k_w=\pi/2$ (Left) and $k_x=0$, $k_z=\pi/2$ (Right). Simulations (Top) qualitatively agree with TB results (Bottom), with black and red/blue labeling bulk and left/right edge modes respectively.}
\label{pgcomsol}
\end{figure}

For concreteness, consider the photonic unit cell of Fig.~\ref{pgcomsol}(a). It consists of eight $12^\circ$ tilted rods A to H with equal eccentricities, grouped into four sub-cells with position labels $(x,y)=(0,0),(0,1),(1,0),(1,1)$. An effective 4D system emerges from synthetic DOFs $k_z,k_w$ which tune the rods major axes according to their sub-cell positions, as manifested from on-site energies $H_{onsite}\propto \cos{(\pi x+k_z)}+ \cos{(\pi y+k_w)}$ (see~\cite{SOM} for the tight-binding model). In particular, the nonsymmorphic pg symmetry $g_y=\{m_y|\tau_x\}$ exists in the 3D planes $k_z=\pm\pi/2$, since rods A to D and rods E to H will have equal major axes respectively and hence respect the nonsymmorphic condition $g_y\psi(x,y)=\psi(x+\frac{a}{2},-y)$.

In this quasi-4D system, line degeneracies i.e.~nodal lines of the $p_x,p_y$ bands emerge, for instance, in the $k_x$-$k_w$ plane at $k_y=0,k_z=\pi/2$ (inset of Fig.~\ref{pgcomsol}(b)). In certain synthetic momentum slices, i.e. $k_w=\pi/2$, the degeneracy becomes a Dirac point, as illustrated by the purple dot in both Fig.~\ref{pgcomsol}(b) (upper panel) and its inset. The crucial role of nonsymmorphic symmetry $g_y$ can be see by tilting rods B,D,F,H i.e. to $30^\circ$, which gaps out the line degeneracies (Fig.~\ref{pgcomsol}(b) lower panel). Such phase transitions crucially rely on elliptical rods that are easily engineered in our photonic setup, but are impractical in material realizations.

Another interesting aspect of our 4D nodal system, as compared to its 3D counterparts, is that topological boundary states can be pumped without merging into the bulk. Unlike ``Drumhead" surface states of 3D systems which do not extend throughout the whole brillouin zone (BZ), our surface states are only bounded by ``slices" of the nodal boundary projection, and can remain separated from the bulk across the whole synthetic parameter period~\cite{kraus2013four,Zilberberg2018}. This is shown in Fig.\ref{pgcomsol}(d) with open boundaries in $y$ (See SOM~\cite{SOM} for more slices).

\noindent\textit{Experimental prospects --}
With current advances in fabrication technology \cite{PhysRevA.92.023849,hsieh2015photon,wagner2016two}, our proposed entangled modes can be experimentally observed relatively simply. We propose a 2D PC (Fig.~\ref{fig:sdefect}b) consisting of a background lattice of rods of radius $0.2a$ and dielectric constant $8$, with $a$ in the decimeter range. Every third rod is replaced by an otherwise identical defect rod of radius determined by Eq.~\ref{eq4}. The spectral flow of the edge modes can be measured when synthetic dimension parameters $k_z,k_w$ are varied over a period. This edge mode exists in the bulk gap, and will appear as a distinct peak in the  microwave transmission spectrum~\cite{Maser1,Maser2,Maser3} when the wavelength is tuned within the bulk gap. With our entangled 4D QH mode, a resonant peak is for instance expected at the wavelength $0.2833 m$, i.e. when $k_z=-1.37$ and $k_w=0$ (Fig. \ref{fig:sdefect}). Alternatively, the edge mode can also be probed via photonic topological boundary pumping with paraxial (out of plane) light~\cite{Zilberberg2018}.

\noindent\textit{Conclusions --}
We have proposed a very versatile and realistic way of realizing analogs of generic multi-component gapped QH and gapless nodal topological states through spatially modulated 2D photonic lattices. Modes around the defect rods are governed by Hofstadter Hamiltonians, and consequently behave like QH wavefunctions. With quasi-periodic modulations coupling the two directions, we obtain hitherto unreported entangled 4D QH modes and nodal lines/points protected by nonsymmorphic symmetry. Such is the versatility of our PC platform that by simply modulating the rod size/shapes, existing PC experiments on "designer" disorder, non-Hermitian~\cite{liang2017}, high-order~\cite{Rechtsman}, interacting~\cite{michael2006} and optomechanical~\cite{peano2015topological} topological/non-topological effects can be naturally generalized to higher dimensions. Applications include one-way optical fibers, with topologically protected chirality in the space of in/out-of plane modes. 


\noindent\textit{Acknowledgements--}
C.H.L. thanks Oded Zilberberg, Mu Sen and Jiangbin Gong for discussions. X.Z. is supported by the National Natural Science Foundation of China (Grant No. 11874431), the National Key R \& D Program of China (Grant No. 2018YFA0306800) and the
Guangdong Science and Technology Innovation Youth Talent Program (Grant No. 2016TQ03X688). Bochen Guan is supported by a scholarship from the Oversea Study Program of Guangzhou Elite Project.
\newpage

\begin{widetext}
\begin{center}
\textbf{\large Supplemental Online Material for \\
``Entangled 4D multi-component topological states from photonic crystal defects" }\\[5pt]
\end{center}

\setcounter{equation}{0}
\setcounter{figure}{0}
\setcounter{table}{0}
\setcounter{section}{0}
\makeatletter
\renewcommand{\theequation}{S\arabic{equation}}
\renewcommand{\thefigure}{S\arabic{figure}}
\renewcommand{\thesection}{S\Roman{section}}
In this Supplemental Material, we give some details on the main text in the following parts:
\begin{itemize}
\item Section A\ref{AppA} shows the dependence of the onsite frequencies of various $s, p$ and $d$-type defect modes with the defect rod radius $r$ and the frequency dispersion for the various orbital modes explicitly.

\item Section B\ref{AppB} presents numerical results on topological phases arising from a $1D$ quasi-periodic lattice of defect rods to supplement the discussion in the main text, which extends to the $4D$ quantum Hall (QH) modes.

\item Section C\ref{AppC} elaborates how two unentangled Hofstadter models form a $4D$ QH system and give rise to a factorizable $2$nd Chern number, distinguished from the entangled case in the main text.

\item Section D\ref{AppD} provides the technical details of calculating the $1$st and the $2$nd Chern number via a Wilson loop approach. 

\item Section E\ref{AppE} derives the relationship between the Berry curvature entanglement eigenvalues $\tilde \lambda_i$ and the wavefunction/state entanglement eigenvalues $\lambda_i$ in detail.

\item Section F\ref{AppF} elucidates why the electromagnetic response of 4D systems with entangled 2nd Chern numbers cannot be decomposed into two $1$st Chern number responses.

\item Section G\ref{AppG} analyzes the stability of the $4D$ QH state by introducing Gaussian random fluctuation to radii of the defect rods and simulating the corresponding energy band structure. 

\item Section H\ref{AppH} illustrates the tight-binding model for 4D nonsymmorphic pg lattice and compares the theoretical results with the simulations by COMSOL Multiphysics.
\end{itemize}
\tableofcontents

\newpage

\section{A. $s,p$ and $d$-type defect modes}\label{AppA}
The shape of a defect mode depend qualitatively on the defect rod radius. For different radii, the defect modes can exhibit $s,p$ or $d$-type symmetries (Fig.~\ref{HigerorbitFIG.pdf}), hence providing a monolithic platform for constructing multi-component topological states without the need for multiple physical ``orbitals''.
\begin{figure}[h]
\centering
\includegraphics[width=.9\linewidth]{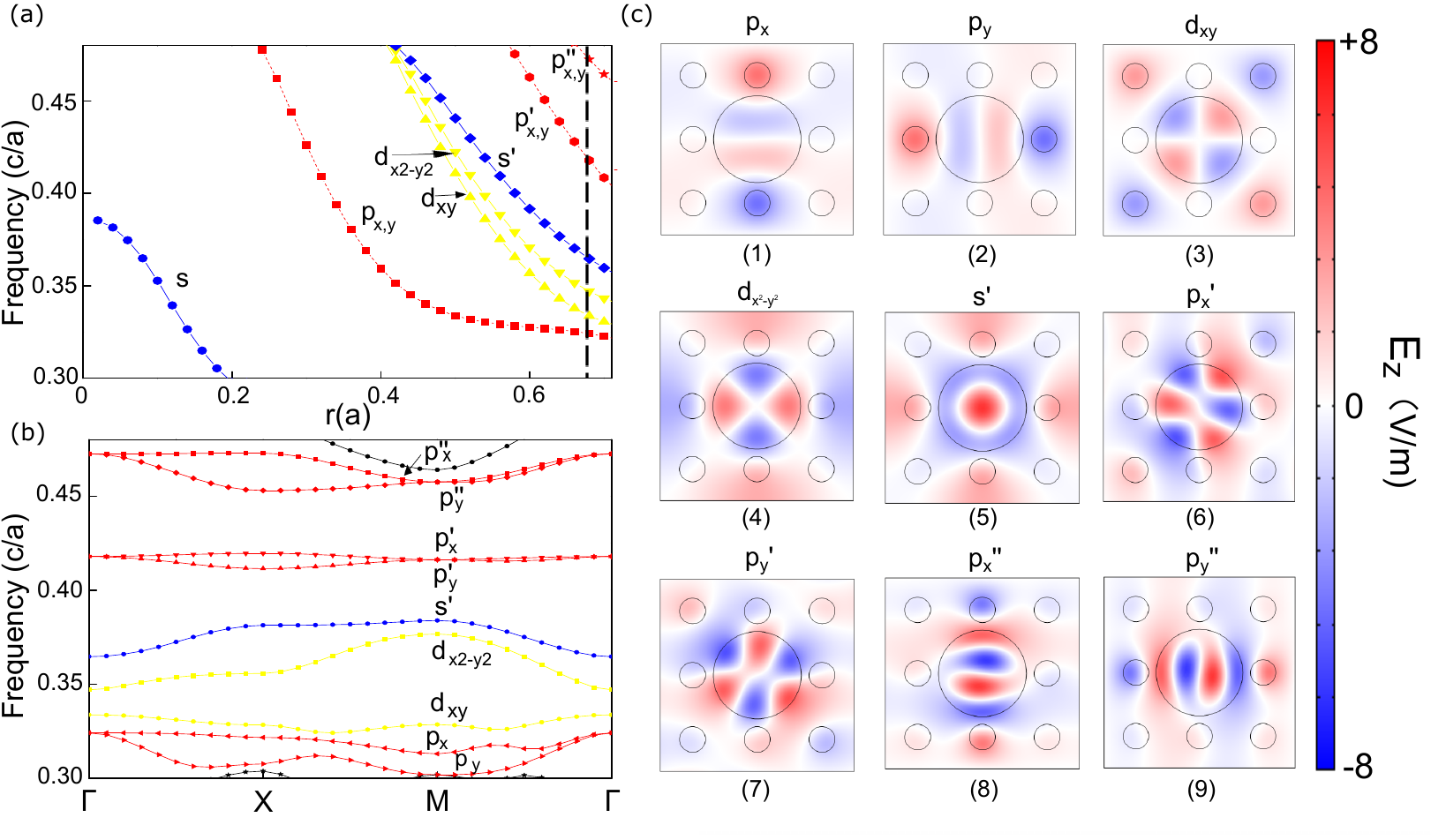}
\caption{(a) Dependence of the onsite frequencies of various $s,p$ and $d$-type defect modes with the defect rod radius $r$. Only $s$-type modes exist for $r<0.2a$, the background rod radius, but progressively more exotic modes appear at larger radii. (b) Frequency dispersion for the various orbital modes within the background bulk gap at an illustrative defect radius of $r=0.68a$. (c) The $E_z$ distributions of the various modes in the photonic crystal structure as Fig.2(b) in the main text shows. The proliferation of these modes lead to interesting multi-component effective Hamiltonians. }
\label{HigerorbitFIG.pdf}
\end{figure}
\section{B. Topology of a 1D lines of defects\label{AppB}}
In this section, we shall present numerical results on topological phases arising from 1D defect lattices to supplement the discussion in the main text, from which more interesting 4D QH analogs are based on. To set the stage for realizing our entangled 4D quantum Hall (QH) modes, we first describe how a \emph{1D} quasi-periodic lattice of defect rods can give rise to topological boundary modes protected by the 1st Chern number of an effective \emph{2D} QH system. This can be achieved extremely simply by spatially modulating a 1D line of defects with radii $r(x)$ according to 
\begin{equation}
r(x)=r_{0}+r_{1} \cos(2\pi b x+k_y),
\label{eq1}
\end{equation}
where $r_0,r_1$ are constants, $b$ is a rational modulation frequency  and $x$ labels the defect sites in the $\hat x$-direction (Fig.~\ref{fig:2a_new}d). If we choose $r_0+r_1<0.2a$, only the simplest $s$ orbital modes are present. To realize $p$-wave modes, we need the defect radii to exceed $0.2a$ instead. $k_y$ controls the phase of the modulation by setting the radius of the first defect rod, and takes the role of a synthetic dimension parameter. The frequency spectrum of the defect modes can be computed by a finite-element Maxwell's equation (FEM) solver, and is presented, for $b=1/4$, in Fig.~\ref{fig:2a_new}a,b for $s$-type defect modes with perodic and open boundary condition (PBC/OBC)  in the $\hat x$-direction. To be more specific, we employed the “Electromagnetic Wave, Frequency Domain” Module in the simulation software COMSOL Multiphysics, adopted MUMPS (Multifrontal Massively Parallel sparse direct Solver) as the eigenvalue solver and a physical-controlled mesh with finer element size during the simulation. Simulated in Fig. \ref{fig:p1d}a,b, the bandstructure (for $b=1/4$) of $p$-wave modes with PBC and OBC consists of 8 bulk bands, twice of the $s$ bulk bands. 

\begin{figure}[H]
\centering
\includegraphics[width=.9\linewidth]{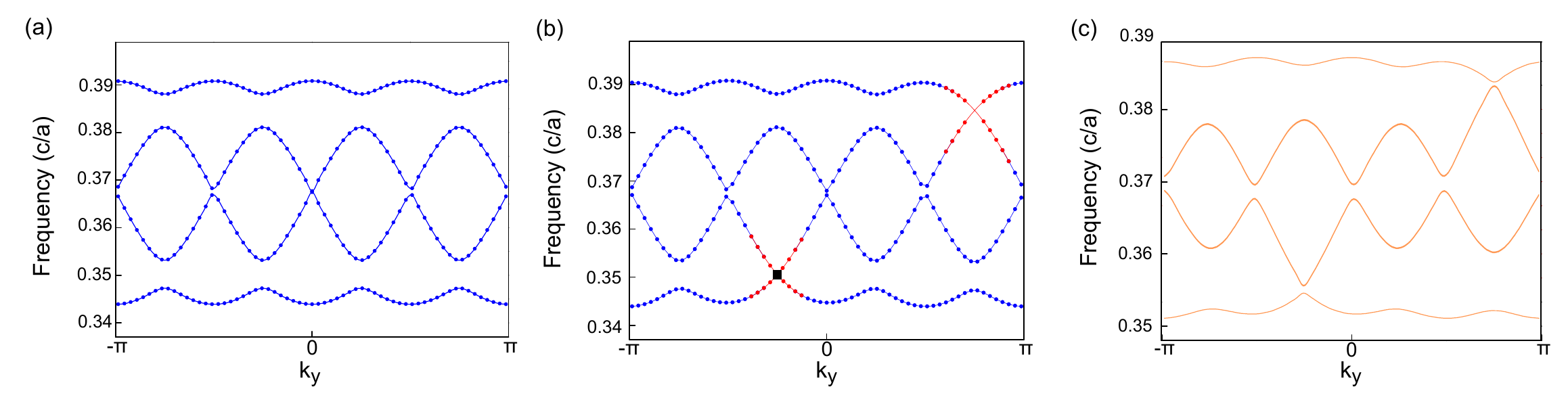}
\caption{ Frequency dispersion results from finite-element method (FEM) calculations with COMSOL Multiphysics. The system consists of a line of defects with radii $r(x)=0.1a+ 0.03a \cos (2\pi b x+k_y)$, $b=1/4$ with (a) periodic and (b) open boundary conditions, i.e. defects terminated by the background lattice in (d). With open boundary condition, the frequency dispersion of $s$-type defect bands exhibit boundary modes (red) that traverse the gap as $k_y$ is varied, indicative of topological pumping by the 1st Chern number. The red regions represent modes localized at the ends of the defect chain. (c) Tight-binding approximation to the OBC system in b) with fitted parameters $\omega_{0}=0.36862(c/a)$, $\lambda=-0.0174096$ and $t=-0.00398(c/a)$, showing very good numerical agreement.  }
\label{fig:2a_new}
\end{figure}

Upon varying $k_y$ over a period, boundary modes in the OBC case are seen to continuously connect the bulk bands. This spectral flow of mid-gap states is signature of topological behavior, and can indeed be understood via the effective TB Hamiltonian for $s$-type orbitals
\begin{equation}
\begin{split}
H_{1D,\text{s-type}}=&\sum_{x}(\omega_0+\lambda\cos(2\pi b x + k_y))|\psi_{x}\rangle \langle \psi_{x}|\\
& +t\sum_x( |\psi_{x}\rangle \langle \psi_{x+1}|+h.c.),
\end{split}
\label{AA}
\end{equation}
which follows from the modulation in the defect radius (Eq.~\ref{eq1}), showing very good numerical agreement with the simulation (Fig.~\ref{fig:2a_new}c). $\omega_0$, $\lambda$ and $t$ are constant parameters corresponding to the $s$-type orbitals. Hence we recover a variation of the 1d Aubry-Andr\'e-Harper (AAH) model~\cite{AAH}, which can be mapped to a 2D integer QH system (Chern lattice) with a flux of $2\pi b$ per plaquette~\cite{kraus2012topological,madsen2013topological}, with perfectly flat Chern bands as $b$ tends to an irrational value~\footnote{Writing $b=p/q$, the bands become increasingly flat as $q$, which controls the magnetic unit cell size, becomes large. Perfect flatness is possible only with infinite $q$ due to topological constraints \cite{chen2014impossibility,lee2016band,read2016compactly}.}. To see that this Hamiltonian describes a 2D Quantum Hall (QH) lattice, we first reinterpret the modes $|\psi_{x}\rangle$ as Landau gauge wavefunctions with momentum quantum number $k_y$. Eq.~\ref{AA} then becomes a hopping Hamiltonian on a rectangular lattice with horizontal and vertical hoppings of magnitudes $t$ and $\lambda/2$. Each upward/downward vertical hopping on column $x$ also involves a phase shift factor $e^{\pm 2\pi b i}$. The net phase shift around a closed plaquette is thus $2\pi b$, corresponding to a net magnetic flux threaded per plaquette, i.e. a QH lattice.

When the flux is rational, i.e. $b=p/q$ where $p$ and $q$ are relatively prime, Eq. \ref{AA} describes a Chern insulator with $q$ bands, each with Chern numbers detailed as in Ref.~\onlinecite{hatsugai1993edge}. If $b$ is irrational, we can approximate $b\approx p/q$ with arbitrarily large $p$ and $q$, with which the bands tend towards perfectly flat 'Landau levels' amidst the signature Hofstadter butterfly spectrum~\footnote{Indeed, an infinite number of bands is required for perfectly flat Chern bands, as can be proven via K-theory~\cite{chen2014impossibility,lee2016band,read2016compactly}. }.

Viewed as a Chern insulator, the midgap states in Fig.~\ref{fig:2a_new}b are just the topological boundary modes corresponding to the bulk chiral anomaly~\cite{qi2008topological}. Physically, they realize the pumping mechanism in Laughlin's spectral flow argument, even though it is the defect modes and not electrons that are pumped.  The number of states pumped over a period in $k_y$ depends on the difference in the combined Chern numbers of the bands on either side of gap.

\begin{figure}[H]
\centering
\includegraphics[width=.85\linewidth]{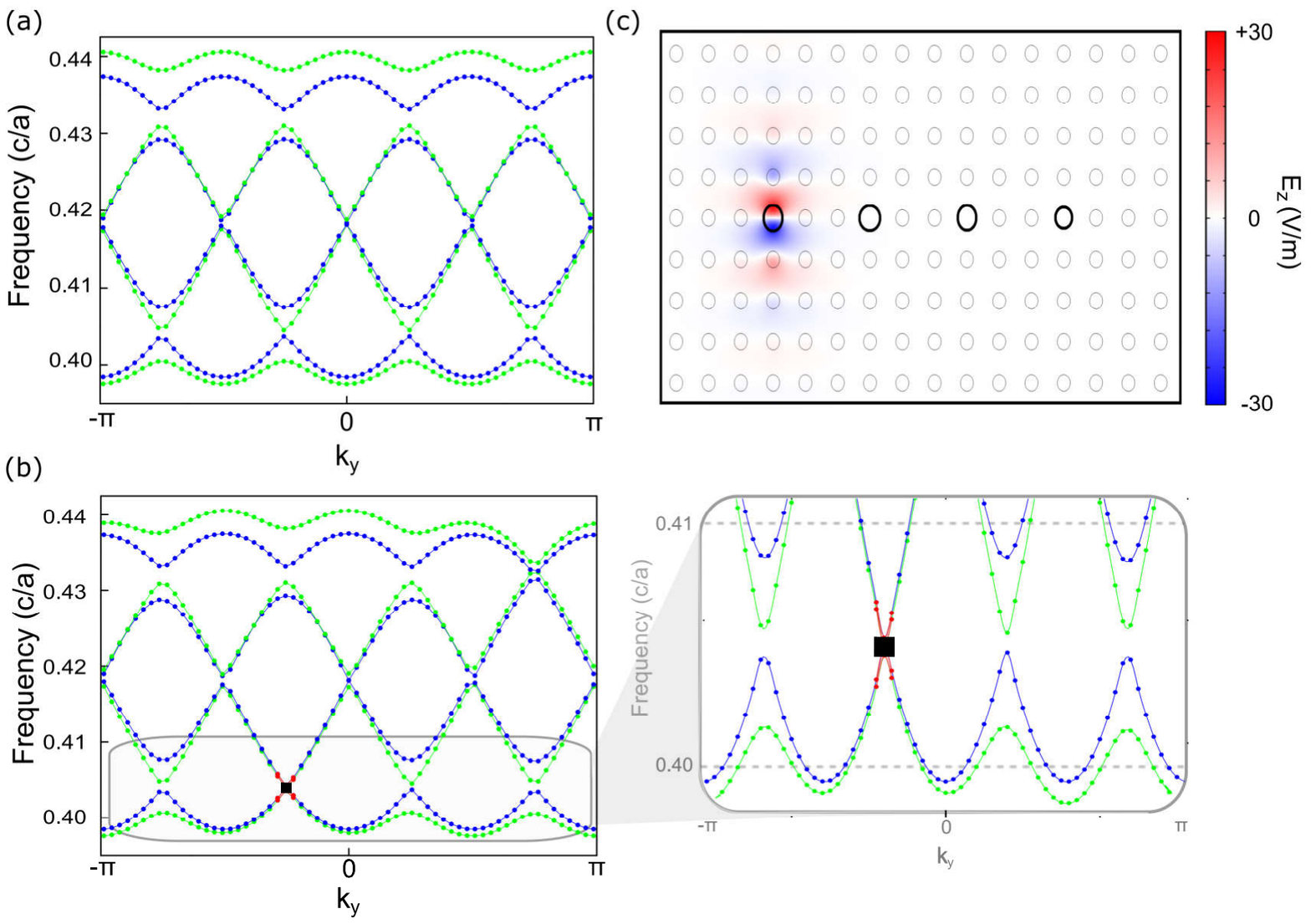}
\caption{(a) The FEM COMSOL frequency bandstructure of $p$-type defect modes for a 1D PC resonator lattice embedded in a background 2D PC, as a function of $k_y$($k_x=0$). The radius of the defect rods obeys Eq.~\ref{eq1} with parameters $r_0=0.3a$, $r_{1}=0.005a$ and $b=p/q=1/4$. The  simulation is performed with a repeated (periodic boundary condition) super-cell having $q=4$ defect rods. Blue(green) lines represent $p_y$ ($p_x$) orbitals, which do not couple to each other. By fitting the band structure with tight binding model in Eq.~\ref{AA1dp}, we obtain $t_{px}=0.00828(c/a)$, $t_{py}=-0.001645(c/a)$, $\lambda=0.01317$ and $\omega_0=0.419515$. (b) Same as in a), but with open boundary conditions defined by a super-cell terminated by regular bulk dielectric rods of radius $0.2a$ on both ends. Note the edge modes traversing the gap in the magnified panel. (c) The $z$-direction electric field strength ($E_z$) distribution of a localized defect boundary mode with $p$-wave symmetry above the first defect. This mode corresponds to the small black box in b), with $k_y=-0.7$  and $\omega=0.4147 (c/a)$.}
\label{fig:p1d}
\end{figure}

With $p$-type orbitals, the effective TB Hamiltonian takes the form
\begin{equation}
\begin{split}
H_{1D,\text{p-type}}=&\sum_{x}(\omega_{0}+\lambda\cos(2\pi b x + k_y))\\&(|px_{x}\rangle \langle px_{x}|+|py_{x}\rangle \langle py_{x}|)\\
+&\sum_{x,x+1}(t_{px} |px_{x}\rangle \langle px_{x+1}|+t_{py} |py_{x}\rangle \langle py_{x+1}|)
\end{split}
\label{AA1dp}
\end{equation}
As shown in Fig. \ref{fig:p1d}a,b, blue(green) lines represent $p_y$ ($p_x$) orbitals, which do not couple to each other. With parameters in Fig. \ref{fig:p1d}, the band gaps of  $p_x$ and $p_y$ orbitals overlap with each other so the system is equivalent two copies of $s$-type defect modes. Time-reversal breaking in the QH (Chern) phase is physically implemented by breaking the lattice translation symmetry. As shown in Fig.~\ref{fig:2a_new}d and \ref{fig:p1d}c, midgap states exist at the edge (ends) of the defect line, and are topologically pumped across the gap as the synthetic dimension parameter $k_y$ cycles over a period~\cite{huang2012entanglement,alexandradinata2014wilson,lee2015free}.

\section{C. Topology of 2D defect lattices\label{AppC}}
A defect lattice with modulations in both directions maps to two copies of 1D AAH models, each corresponding to a 2D Hofstadter QH system with a synthetic dimension. Together, they correspond to a 4D QH system described by a 2nd Chern number. In the main text, we have discussed the case where the two Hofstadter models are entangled. Below, we shall elaborate more on the unentangled case, where the 2nd Chern number factorizes into two 1st Chern numbers viz. Eq. 6 of the main text.

The defect radius of the $(x,y)$-th defect is given by
\begin{equation}
r(x,y)=r_{0}+r_{1} \cos(2\pi b x+k_z) +r_{2} \cos(2\pi b y+k_w),
\label{eq3}
\end{equation}
where $b$, as before, is the flux.

\begin{figure}[h]
\centering
\includegraphics[width=.8\linewidth]{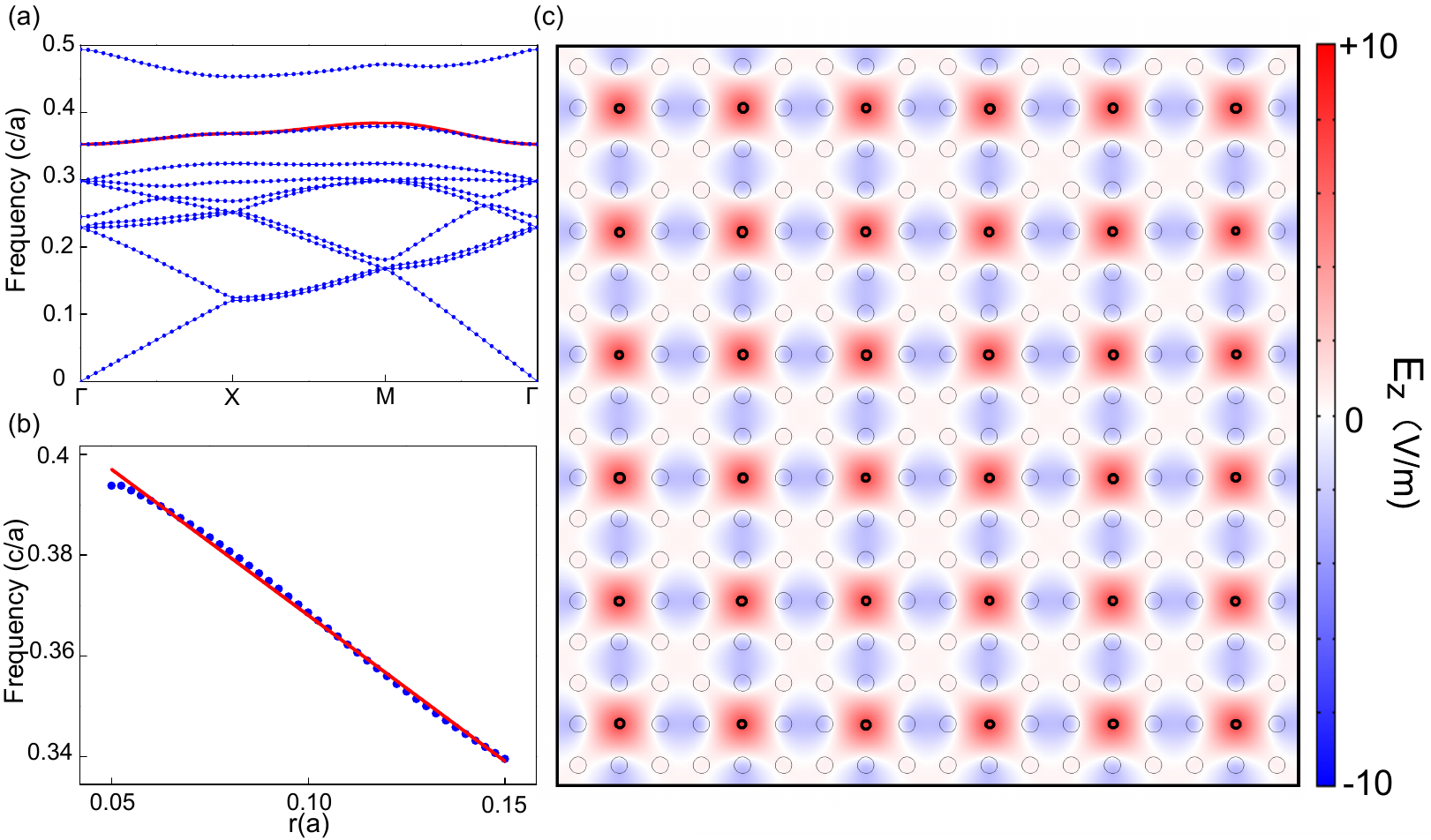}
\caption{ (a) Frequency bands for the TM modes of an array of \emph{identical} thin defects rods of radii $r=0.1a$ and relative dielectric constant 8 embedded within a background array of rods of radii $0.2a$. The energy bands depicted by the blue points are calculated with COMSOL Multiphysics from finite-element method (FEM), including both the bulk and defect modes.  Moreover the red line shows the defect mode only from the TB model. The correspondence between the defect mode represented by blue dots and the red line indicates that our model fits well with the simulation result.Numerical fitting yields $\omega_0=0.36862(c/a)$ and $t=-0.00398(c/a)$. (b) The blue dots shows the monotonic dependence on defect radius of the frequency $\omega_0$ of the $s$-type defect state at the $X$-point calculated by FEM, which is well-approximated by the TB model with $\omega_0=-0.58032(c/a^{2})r+0.42608(c/a)$ as the red line shows. (c) $E_z$ field distribution of the well-localized $s$ defect states. }
 \label{fig:1}
 \end{figure}
\subsection{Factorizable 2nd Chern number from 2D defect lattice with $s$-wave modes\label{AppC1}}
Picking small $r_0,r_1$ and $r_2$ in Eq.~\ref{eq3} such that $r(x,y)<0.2a$, the radius of the background rods, only $s$-type modes appear and the effective Hamiltonian takes the form
\begin{equation}
\begin{split}
H_{2D,s-type}&=\sum_{x,y}(\omega_0+\lambda \cos(2\pi bx+k_z))\left|\psi_{(x,y)}\right\rangle \left\langle \psi_{(x,y)}\right|\\
&+\sum_{x,y}(\lambda \cos(2\pi by+k_w))\left|\psi_{(x,y)}\right\rangle \left\langle \psi_{(x,y)}\right|\\
&+\sum_{x,y} t \left|\psi_{(x,y)}\right\rangle \left\langle \psi_{(x+ 1,y)}\right|+\sum_{x,y} t \left|\psi_{(x,y)}\right\rangle \left\langle \psi_{(x,y+ 1)}\right| +h.c.\\
\end{split}
\label{TB2}
\end{equation}
Further hopping terms are negligible due to the locality of the $s$-type modes, whose properties are elaborated in Fig.~\ref{fig:1}. Frequency bands of the $s$ defect states of our 2D defect lattice modulated according to Eq.~\ref{eq3} with $r_{0}=0.1a$, $r_{1}=r_{2}=0.025a$ and $b=1/4$ are shown in  Fig.~\ref{fig:3}.

\begin{figure}[H]
\centering
\includegraphics[width=.9\linewidth]{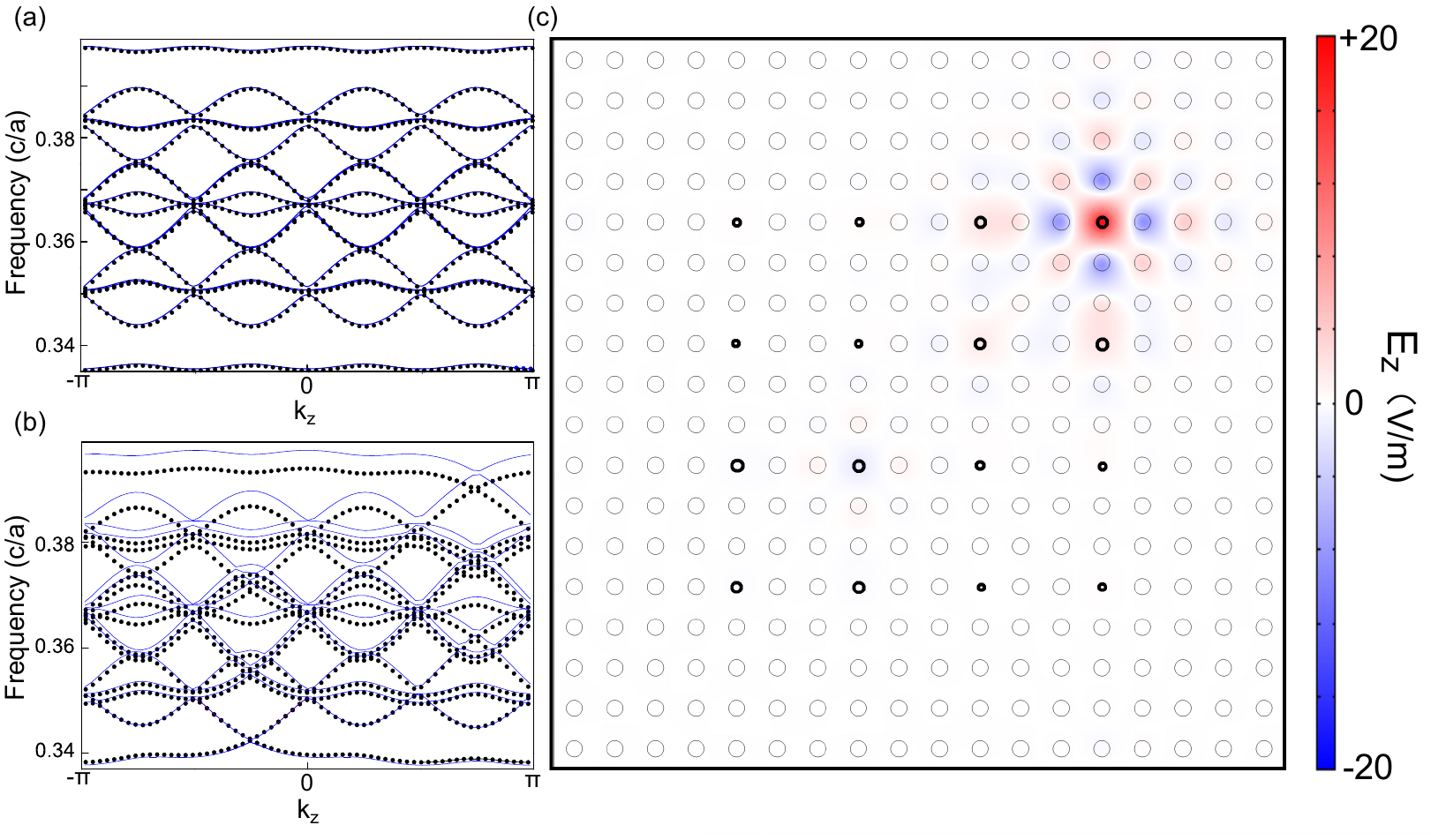}
\caption{(a) and (b): Frequency bands of the $s$ defect states of our 2D defect lattice modulated according to Eq.~\ref{eq3} with $r_{0}=0.1a$, $r_{1}=r_{2}=0.025a$ and $b=1/4$, for periodic and open boundary conditions respectively. To implement open boundary condition, we surround the defect lattice with a set of background pillars as (c) shows, the same as the 1D situation. The radii of the background rods are all $0.2a$. They are plotted as a function of synthetic dimension parameter $k_z$, with $k_w$,$k_x$,$k_y$ set to $0$. Black dots represent FEM COMSOL simulation data, which lie closely on the blue fitted TB curves from Eq.~\ref{TB2} with $\omega_0=0.36862(c/a)$ and $\lambda=-0.014508$ and $t=-0.00398(c/a)$. (c) $E_z$ distribution of a corner defect boundary state at $k_z=-0.8$ and $\omega=0.3423(c/a)$, which lies within the bottom bulk gap in b). This 2D corner defect mode is entirely factorizable into a tensor product of two 1D defect modes.}
\label{fig:3}
\end{figure}

\subsection{Zero 2nd Chern number from 2D defect lattice with $p$-wave modes\label{AppC2}}

We then construct this defect lattice in Fig. \ref{fig:9} using a similar setup as Fig. \ref{fig:3}, but with thicker defect rods that admit $p$-type modes, i.e. with radii $r(x,y)>0.2a$, as given by Eq.~\ref{eq3}. The corresponding effective Hamiltonian is of the form
\begin{equation}
\begin{split}
H_{2D,p-type}&=\sum_{x,y}(\omega_0+\lambda \cos(2\pi bx+k_z))\left|px_{(x,y)}\right\rangle \left\langle px_{(x,y)}\right|\\
&+\sum_{x,y}(\omega_0+\lambda \cos(2\pi by+k_w))\left|py_{(x,y)}\right\rangle \left\langle py_{(x,y)}\right|\\
&+\sum_{x,y}t_{\sigma}(\left|px_{(x,y)}\right\rangle \left\langle px_{(x+1,y)}\right|+\left|py_{(x,y)}\right\rangle \left\langle py_{(x,y+1)}\right|)\\
&+\sum_{x,y} t_{\pi}( \left|px_{(x,y)}\right\rangle \left\langle px_{(x,y+1)}\right|+\left|py_{(x,y)}\right\rangle \left\langle py_{(x+1,y)}\right|)+h.c.
\end{split}
\label{tbp1}
\end{equation}
Further hoppings are negligible due to the locality of the $p$-type modes, whose properties are elaborated in Fig.~\ref{fig:8}.

Since there are no diagonal couplings and the eigenmodes $\left|px\right\rangle$ and $\left|py\right\rangle$ do not couple to each other, Eq.~\ref{tbp1} immediately decomposes into the sum of two independent tensor products. This Hamiltonian possess a zero 2nd Chern number because each of its decoupled $|px\rangle$ and $|py\rangle$ subsystems possess only one nonvanishing Berry curvature. For the $|px\rangle$/$|py\rangle$ subsystem, the only non-vanishing berry curvature is $F_{xz}$/$F_{yw}$. As such, only the 1st Chern numbers $C_1^{xz}$ and $C_1^{yw}$ are nonzero.
\begin{figure}[H]
\centering
\includegraphics[width=.9\linewidth]{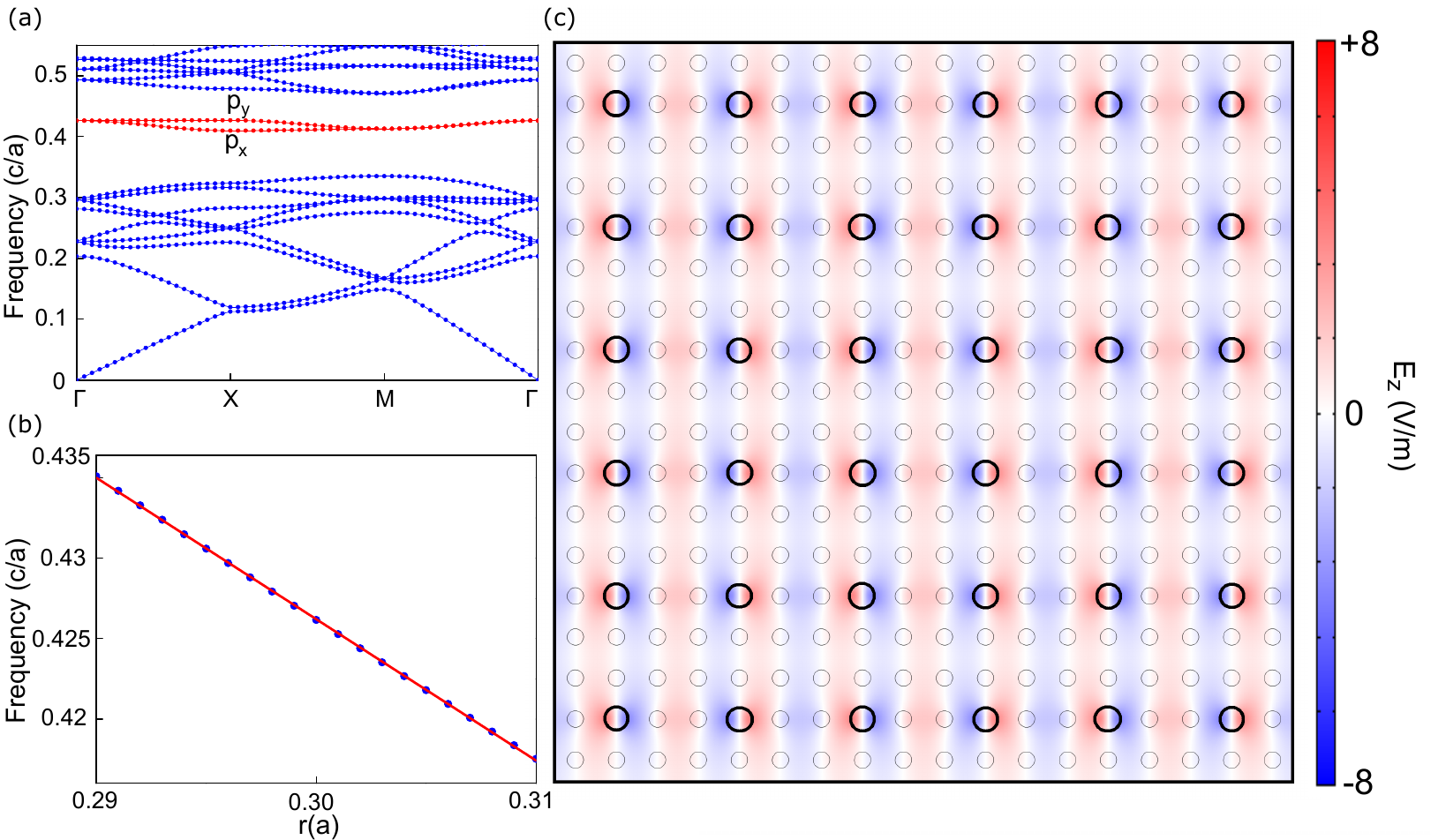}
\caption{ (a) Frequency spectrum of $p$-type defect modes (red) within the bulk gap of modes from the background lattice (blue). Defect rods with radii $r=0.3a$ replace every third background rod of radius $0.2a$. (b) The almost exactly linear dependence of the onsite energy of the defect modes with defect radius $r$ at the $\Gamma$ point. Curve-fitting yields $\omega=0.68961 (c/a)-0.878(c/a^2) r$. (c) The $z$-component electric field strength ($E_z$) distribution of $p_y$ defect modes at the translation-invariant $\Gamma$ point. They collectively give rise to a tight-binding Hamiltonian with numerically determined parameters $\omega_0=0.419515(c/a)$, $t_{\sigma}=0.00828(c/a)$ and $t_{\pi}=-0.001645(c/a)$. }
\label{fig:8}
\end{figure}

\begin{figure}[H]
\centering
\includegraphics[width=.9\linewidth]{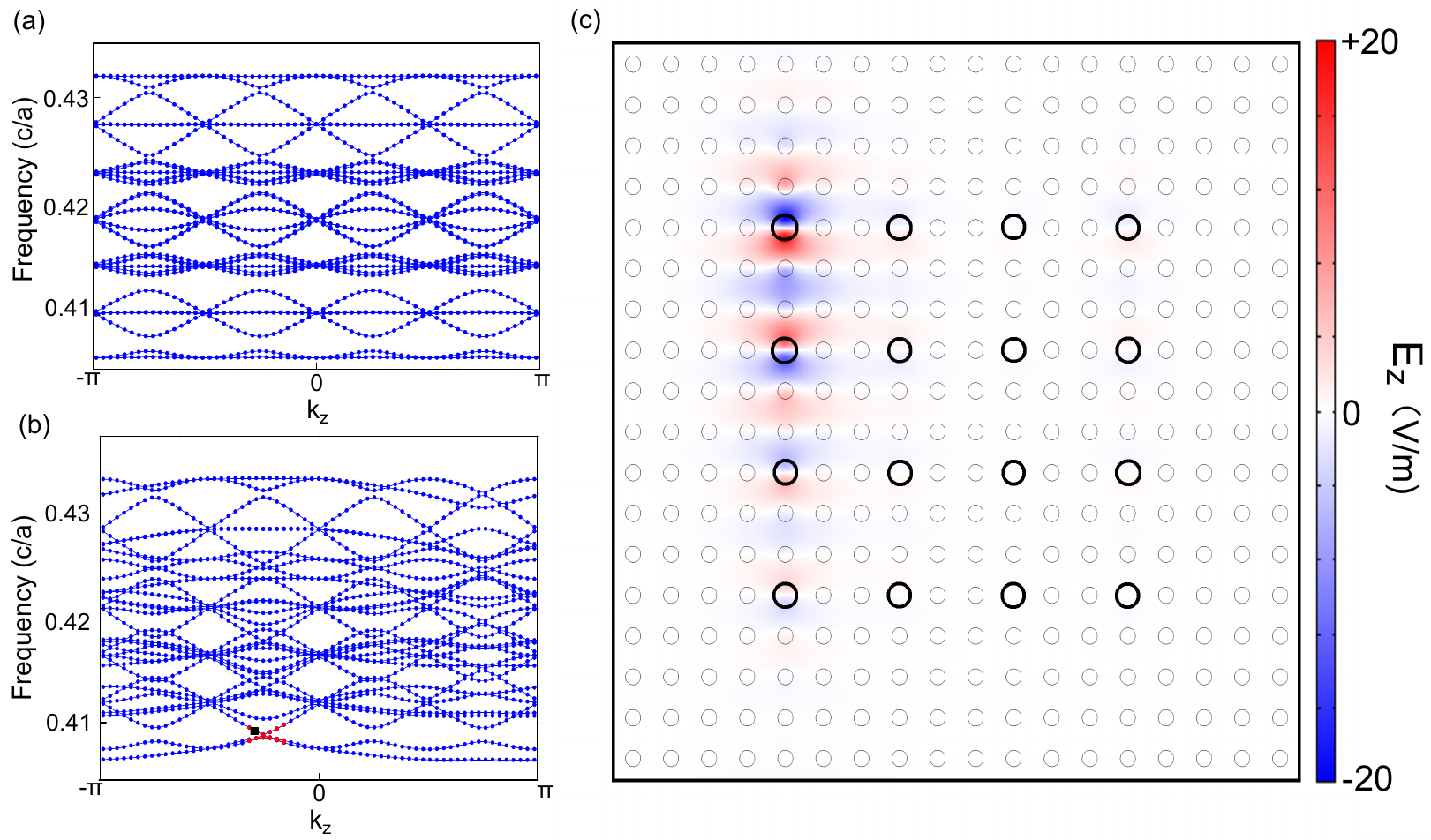}
\caption{(a) and (b): Frequency bands of the defect states of our 2D PC defect resonator lattice with $p$-wave modes, computed with periodic and open boundary conditions respectively. They are plotted as a function of synthetic dimension parameter $k_z$, with $k_w$ set to $0$. Black dots represent FEM COMSOL simulation data, which lie closely on the blue fitted TB curves from Eq.~\ref{tbp1} with $t_{\sigma}=0.00828(c/a)$, $t_{\pi}=-0.001645(c/a)$, $\lambda=0.00439$ and $\omega_0=0.419515$. The radius of the defect rods obeys Eq.\ref{eq3} with parameters $r_{0}=0.3a$, $r_{1}=r_{2}=0.005a$ and $b=\frac{1}{4}$. (c) The electromagnetic field strength $E_z$ of the edge mode (black box of b)) at $k_z=-0.8$, $k_w=0$ and $\omega=0.4086 (c/a)$. The asymmetric localization is due to the asymmetry of its position in the BZ.}
\label{fig:9}
\end{figure}

\section{D. Chern number calculation\label{AppD}}

The first Chern number were computed from the eigenmodes via a Wilson loop approach~\cite{fukui2005chern,PhysRevLett.115.195303}. We first construct a 2D BZ out of a 1D TB model with one existing synthetic dimension $k_\mu$ by threading a flux $t\rightarrow te^{ik_\nu}$ in the real-space direction $x_\nu$. Next we discretize our BZ $[0,\frac{2 \pi}{q}]\times [0,\pi] $ with $N_b=50$ displacement steps in each direction, such that each infinitesimal displacement magnitude is $\epsilon=\frac{2 \pi}{q N_b}$. Each step then gives an infinitesimal abelian Wilson line operator
\begin{equation}
U_{\mu}=\frac{\langle u(\vec{k})|u(\vec{k}+\epsilon \vec{e}_{\mu})\rangle}{|\langle u(\vec{k})|u(\vec{k}+\epsilon \vec{e}_{\mu})\rangle |}
\end{equation}
for each eigenmode $|u(k)\rangle$. If the eigenmodes are not separated by gaps, a \emph{non-abelian} infinitesimal Wilson line operator is given by

\begin{equation}
U_{\mu}=\frac{\begin{bmatrix}
\langle \psi_{a} (\vec{k})|\psi_{a} (\vec{k}+\epsilon \vec{e}_{\mu}) \rangle &\langle \psi_{a} (\vec{k})|\psi_{b} (\vec{k}+\epsilon \vec{e}_{\mu}) \rangle \\
\langle \psi_{b} (\vec{k})|\psi_{a} (\vec{k}+\epsilon \vec{e}_{\mu}) \rangle & \langle \psi_{b} (\vec{k})|\psi_{b} (\vec{k}+\epsilon \vec{e}_{\mu}) \rangle \\	
\end{bmatrix} }{\text{Det}\begin{bmatrix}
\langle \psi_{a} (\vec{k})|\psi_{a} (\vec{k}+\epsilon \vec{e}_{\mu}) \rangle &\langle \psi_{a} (\vec{k})|\psi_{b} (\vec{k}+\epsilon \vec{e}_{\mu}) \rangle \\
\langle \psi_{b} (\vec{k})|\psi_{a} (\vec{k}+\epsilon \vec{e}_{\mu}) \rangle & \langle \psi_{b} (\vec{k})|\psi_{b} (\vec{k}+\epsilon \vec{e}_{\mu}), \rangle \\	
\end{bmatrix}}
\end{equation}
shown here for two modes $a$ and $b$. In both the abelian and non-abelian cases, the Berry curvature is then given by
\begin{equation}
F_{\mu \nu}=\frac{1}{2 \pi i} log\frac{U_{\mu}(\vec{k}) U_{\nu}(\vec{k}+\epsilon \vec{e}_{\mu})}{ U_{\mu}(\vec{k}+\epsilon \vec{e}_{\nu}) U_{\nu}(\vec{k})}
\end{equation}
which integrates to the 1st Chern number
\begin{equation}
C_1=\frac1{2\pi}\int F_{xy} d^2\vec k
\end{equation}

The 2nd Chern number is easily expressed in terms of the three unique 1st Chern numbers via
\begin{equation}
C_2=\frac1{(2\pi)^2}\int F_{zx}F_{yw}+F_{yz}F_{xw}+F_{xy}F_{zw} d^4\vec k
\end{equation}


\begin{figure}[H]
\centering
\includegraphics[width=.5\linewidth]{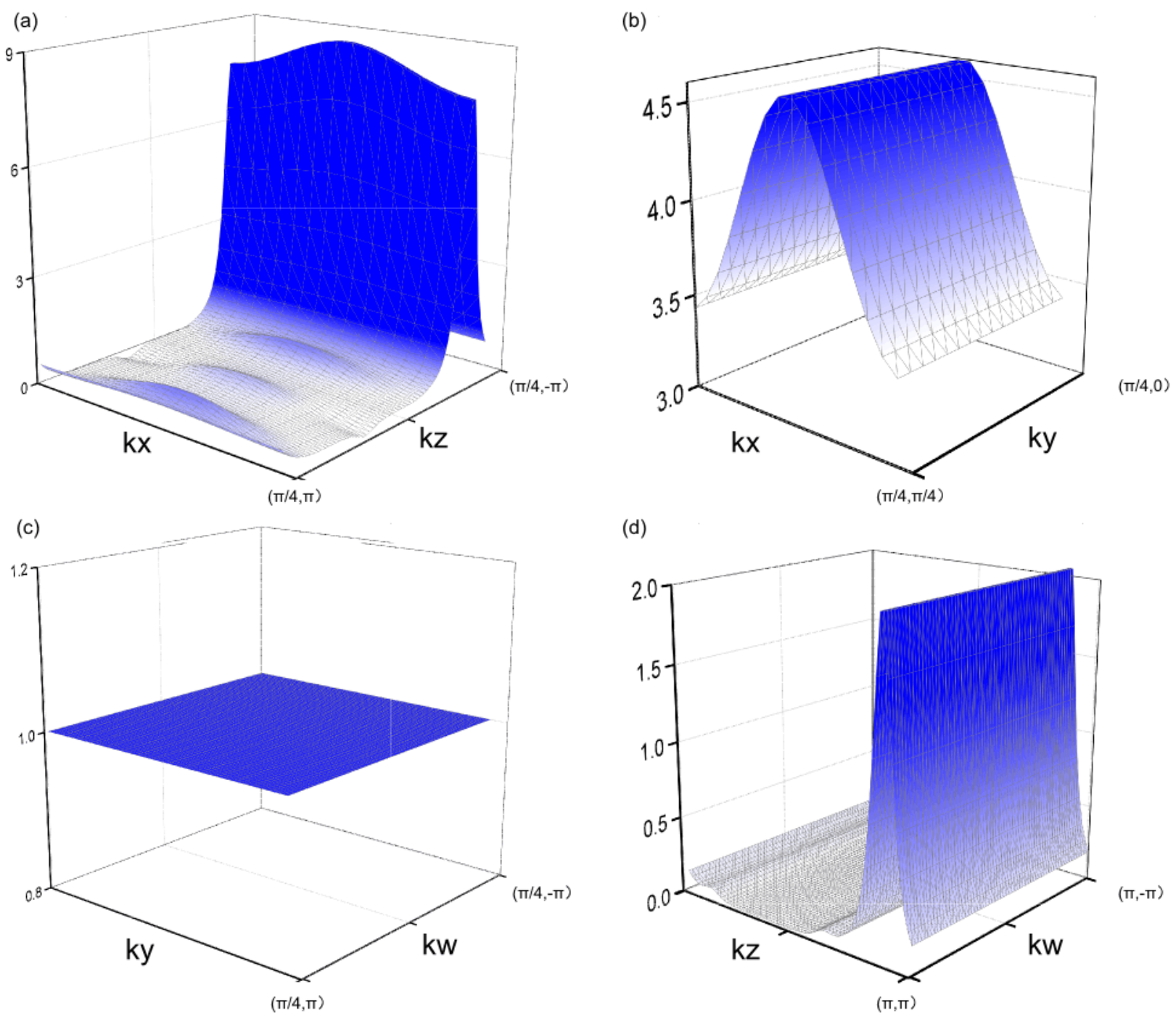}
\caption{Numerically computed Berry curvature of the unentangled 4D QH model of Fig.~3. As a demonstration, plotted are the integrals of $F_{xz}$: (a) $\int F_{xz} dk_{y} dk_{w}$ (b) $\int F_{xz} dk_{z} dk_{w}$ (c) $\int F_{xz} dk_{x} dk_{z}$ (d) $\int F_{xz} dk_{x} dk_{y}$. We obtain $C_2=\frac1{(2\pi)^2}\int F_{zx}F_{yw}+F_{yz}F_{xw}+F_{xy}F_{zw}~d^4\vec k =-1+0+0=-1$.   }
\label{fig:productive}
\end{figure}
\begin{figure}[H]
\centering
\includegraphics[width=.5\linewidth]{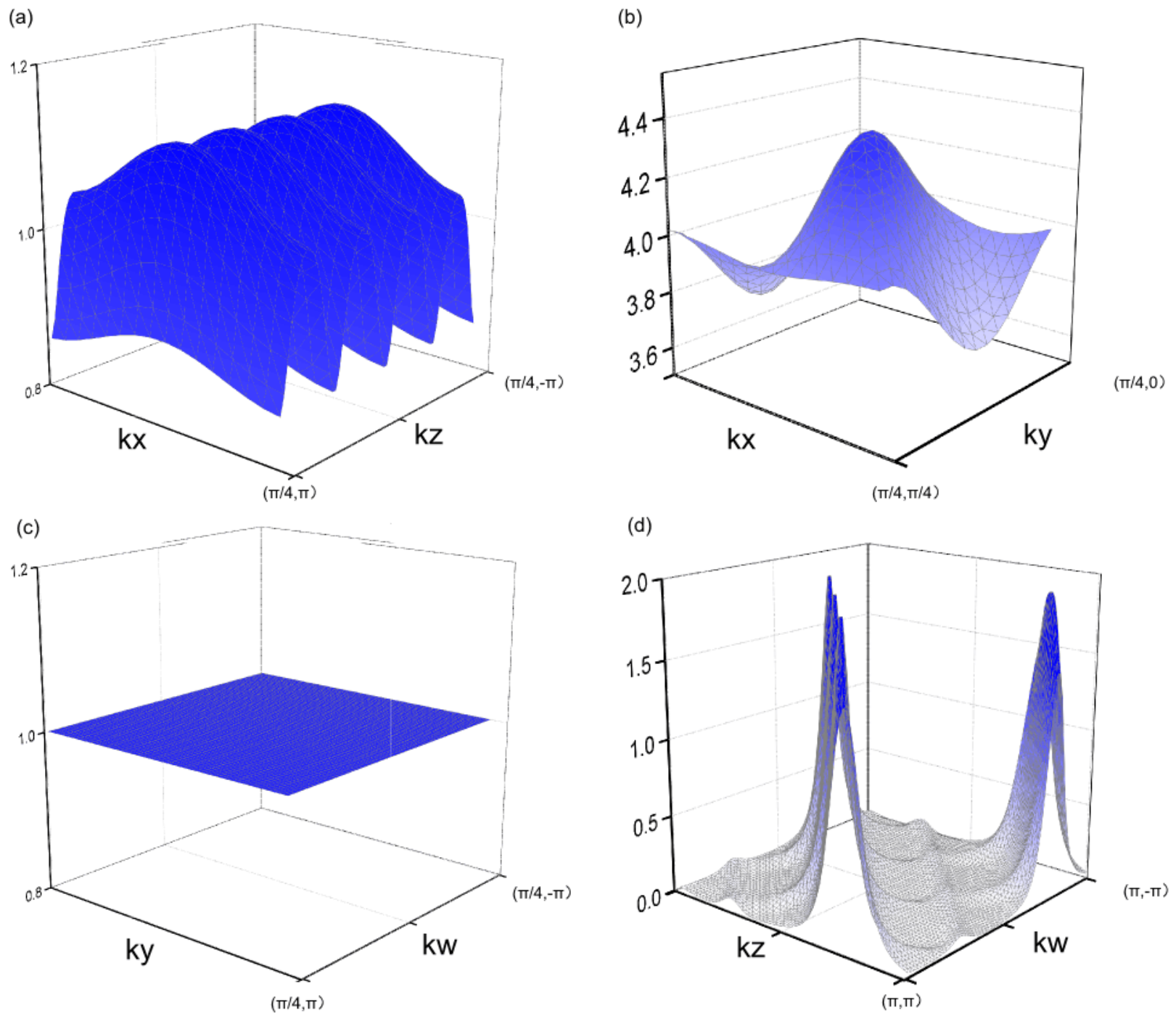}
\caption{Numerically computed non-abelian Berry curvature of the entangled 4D QH model of Fig.~4 of the main text. Plotted are also the integrals of $F_{xz}$: (a) $\int F_{xz} dk_{y} dk_{w}$ (b) $\int F_{xz} dk_{z} dk_{w}$ (c) $\int F_{xz} dk_{x} dk_{z}$ (d) $\int F_{xz} dk_{x} dk_{y}$. We obtain $C_2=\frac1{(2\pi)^2}\int F_{zx}F_{yw}+F_{yz}F_{xw}+F_{xy}F_{zw}~d^4\vec k =1+1+0=2$.  }
\label{fig:nonproductive}
\end{figure}
\section{E. Relationship between wavefunction and Berry curvature entanglement\label{AppE}}
\label{app:entanglement}

We show explicitly how the Berry curvature entanglement eigenvalues $\tilde \lambda_i$ are related to the wavefunction/state entanglement eigenvalues $\lambda_i$. We shall specialize to abelian Berry curvature; the non-abelian case with more than one eigenmode generalizes straightforwardly.

For clarity of notation, we shall switch the momentum variable notation $(k_x,k_y,k_z,k_w)\rightarrow (q_1,p_1,q_2,p_2)$, such that in the case of the 4D QH state being a tensor product of two 2D QH states (labeled $1$ and $2$), the $p_1,q_1$ and $p_2,q_2$ degrees of freedom (DOFs) decouple.

Performing an entanglement cut across the $p_1,q_1$ and $p_2,q_2$ DOFs entails the SVD decomposition of a generic state $\phi$ into
\begin{equation}
\phi=\sum_i\lambda_i \,\phi_1^i\otimes \phi_2^i
\end{equation}
where contributions $\phi^1_i$, $\phi^2_i$ depend only on DOFs $p_1,q_1$ and $p_2,q_2$ respectively. The entanglement eigenvalues are given by $\lambda_1\geq \lambda_2\geq \lambda_3...$, which are normalized according to $\sum_i\lambda_i=1$. In the case of a tensor product 4D state, we have $\lambda_1=1$ and all other $\lambda_{i\geq 2}=0$.

The 4D second Chern number $C_2$ is given in terms of the 2D Berry curvatures $F_{\mu\nu}$ via $C_2=\frac1{8\pi^2}\int  F\wedge F\, d^4 k$, with
\begin{eqnarray}
\frac1{2}F\wedge F &=& F_{q_1p_1}F_{q_2p_2}-F_{q_1q_2}F_{p_1p_2}+F_{q_1p_2}F_{p_1q_2}\notag\\
&=&\sum_{ijkl}\lambda_i\lambda_j\lambda_k\lambda_l \times\notag\\
&&\;\,[(\langle\partial_{q_1}\phi^i_1\otimes \phi_2^i|\partial_{p_1}\phi^j_1\otimes \phi^j_2\rangle-\langle\partial_{p_1}\phi^i_1\otimes \phi_2^i|\partial_{q_1}\phi^j_1\otimes \phi^j_2\rangle)(\langle\phi^k_1\otimes \partial_{q_2}\phi_2^k|\phi^l_1\otimes \partial_{p_2}\phi^l_2\rangle-\langle\phi^k_1\otimes \partial_{p_2}\phi_2^k|\phi^l_1\otimes \partial_{q_2}\phi^l_2\rangle)\notag\\
&& -(\langle\partial_{q_1}\phi^i_1\otimes \phi_2^i|\phi^j_1\otimes \partial_{q_2}\phi^j_2\rangle-\langle\phi^i_1\otimes \partial_{q_2}\phi_2^i|\partial_{q_1}\phi^j_1\otimes \phi^j_2\rangle)(\langle\partial_{p_1}\phi^k_1\otimes \phi_2^k|\phi^l_1\otimes \partial_{p_2}\phi^l_2\rangle-\langle\phi^k_1\otimes \partial_{p_2}\phi_2^k|\partial_{p_1}\phi^l_1\otimes \phi^l_2\rangle)\notag\\
&&+(\langle\partial_{q_1}\phi^i_1\otimes \phi_2^i|\phi^j_1\otimes \partial_{p_2}\phi^j_2\rangle-\langle\phi^i_1\otimes \partial_{p_2}\phi_2^i|\partial_{q_1}\phi^j_1\otimes \phi^j_2\rangle)(\langle\partial_{p_1}\phi^k_1\otimes \phi_2^k|\phi^l_1\otimes \partial_{q_2}\phi^l_2\rangle-\langle\phi^k_1\otimes \partial_{q_2}\phi_2^k|\partial_{p_1}\phi^l_1\otimes \phi^l_2\rangle)]\notag\\
&=& \sum_{ijkl}\lambda_i\lambda_j\lambda_k\lambda_l \times [F^{ij}_{q_1p_1}\langle \phi_2^i|\phi_2^j\rangle\langle \phi_1^k|\phi_1^l\rangle  F^{kl}_{q_2p_2}  -([A^{ji}_{q_1}]^*A^{ij}_{q_2}-[A^{ji}_{q_2}]^*A^{ij}_{q_1})([A^{lk}_{p_1}]^*A^{kl}_{p_2}-[A^{lk}_{p_2}]^*A^{kl}_{p_1})\notag\\
&& +([A^{ji}_{q_1}]^*A^{ij}_{p_2}-[A^{ji}_{p_2}]^*A^{ij}_{q_1})([(A^{lk}_{p_1}]^*A^{kl}_{q_2}-[A^{lk}_{q_2}]^*A^{kl}_{p_1})]\notag\\
&=& \sum_{ijkl}\lambda_i\lambda_j\lambda_k\lambda_l \times (F^{ij}_{q_1p_1}\langle \phi_1^k|\phi_1^l\rangle)\times(F^{kl}_{q_2p_2}\langle \phi_2^i|\phi_2^j\rangle  )\notag\\
&&+ \sum_{ijkl}\lambda_i\lambda_j\lambda_k\lambda_l \times [A_{q_1}^{ji}]^*[A_{p_1}^{lk}]^*\times(A_{p_2}^{ij}A_{q_2}^{kl}-A_{q_2}^{ij}A_{p_2}^{kl})+c.c\notag\\
&&+\sum_{ijkl}\lambda_i\lambda_j\lambda_k\lambda_l \times  [A_{p_1}^{lk}]^*A_{q_1}^{ij}\times([A_{q_2}^{ji}]^*A_{p_2}^{kl}-[A_{p_2}^{ji}]^*A_{q_2}^{kl})+c.c.
\label{C2entangle}
\end{eqnarray}
where $F^{ij}_{\mu\nu}=\langle\partial_\mu \phi^i|\partial_\nu\phi^j\rangle-\langle\partial_\nu \phi^i|\partial_\mu\phi^j\rangle$ is the $(i,j)$-th Berry curvature matrix element contribution, and $A^{ij}_\mu=\langle \phi^i|\partial_\nu\phi^j\rangle$ is its corresponding Berry connection element. There is no ambiguity in whether $\phi^i_1$ or $\phi^i_2$ is used, because each momentum derivative is nonzero with respect to only one of them. Note that the above matrix elements are \emph{not} elements of a non-abelian Berry curvature/connection, because they represent contributions to one single state.

Equation \ref{C2entangle} decomposes $F\wedge F$ into a linear combination of terms, each which is a product of a term depending only on DOFs $q_1,p_1$, and another term depending on DOFs $q_2,p_2$. These terms are weighted by products of the $\lambda_i$, i.e. $\tilde\lambda\propto \lambda_i\lambda_j\lambda_k\lambda_l$, which measure how much the 4D wavefunction is entangled across sectors spanned by momenta $(q_1,p_1)$ and $(q_2,p_2)$. Eq. \ref{C2entangle} illustrates the geometric and entanglement origins of the results of Fig. 3c of the main text.

\section{F. Entangled and unentangled 4D responses through semiclassical wavepacket pumping\label{AppF}}


Here we elucidate why the electromagnetic response of 4D systems with entangled 2nd Chern numbers cannot be understood as the composition of two 1st Chern number responses. We first relate this response to 1D and then 2D semiclassical wavepacket and Wannier center pumpings, and compare the entangled and unentangled cases.

\subsection{Spectral flow for 2D QH\label{AppF2}}
\subsubsection{2D wavepacket response}

In a 2D system governed by the 1st Chern number, consider the motion under the influence of the Berry curvature $F$ of a Bloch wavepacket centered around momentum $\vec k$ with real-space coordinate $X_{\vec k}$. Semiclassical theory tells us that $X_{\vec k}$ responds to an external electric field $E_s$ according to $\dot X_{\vec k}=-\dot  k \times F$, where $ k_s \rightarrow k_s + E_st$ from minimal coupling. Hence, we have
\begin{equation}
X_{\vec k}^s(t)=(k_s+E_st)F_{xs}= X^s(0)+F_{xs}E_st
\end{equation}
where the superscript $s$ has been added to emphasize that the time dependence of $X$ is due to an electric field in the direction $\hat s$. This is just the Hall response of wavepackets.

\subsubsection{2D Wannier center response}

Next, we relate this wavepacket response to Wannier center response. For a 2D QH system in the $x$-$s$ plane, the spectral flow of Wannier function centers is directly related to the Hall pumping of its constituent wavepackets as follows. We write the time-dependent Wannier center as $X(t)$, which is equal to the polarization $P(k_s)$ upon the minimal coupling $k_s\rightarrow k_s + E_st$:
\begin{eqnarray}
X(t)&=&  P_X(k_s+E_st)\notag\\
&= & \frac1{2\pi}\int_0^{2\pi}\int_0^{k_s+E_st} F_{xs}(k'_x,k'_s)dk'_sdk'_x \notag\\
&\sim & (k_s+E_st)\frac{1}{(2\pi)^2}\int_0^{2\pi}\int_0^{2\pi}F_{xs}(k'_x,k'_s)dk'_sdk'_x \notag\\
&=& (k_s+E_st)\int_{[0,2\pi]^2}\frac{F_{xs}}{(2\pi)^2}d^2\vec k' \notag\\
&=& X(0)+\frac{C_1^{xs}}{2\pi}E_st\notag\\
&=&\frac1{(2\pi)^2}\int_{[0,2\pi]^2}X^s_{\vec k'}(t)d^2\vec k'
\label{2D}
\end{eqnarray}
In the 2nd line, I have used the standard expression for the Wannier polarization $P_X$ of $X(t)$ due to $k_s$, which for macroscopic displacements can be linearly approximated as in line 3. 
In the second last line, I have equivalently expressed $X(t)-X(0)$ in terms of the Chern number $C_1^{xs}$.

Not surprisingly, the Wannier center $X(t)$ is given, in the last line, by the integral of the centers of all wavepackets in the BZ.

\subsection{Spectral flow for 4D QH\label{AppF3}}
\subsubsection{4D wavepacket response}

In 4D, there exists nontrivial pumping contributions up to 2nd order in the perturbing fields $E$ and $B$. The key picture is to interpret the 2nd order response as a magnetoelectric response consisting of a magnetic field Hall response of a Berry curvature Hall response.

Consider first an \emph{unentangled} 4D QH system spanning coordinates $x$-$y$-$s$-$u$. We shall show that the averaged wavepacket position $X(t)$ possess response properties exactly given by the 1st and 2nd Chern numbers. Qualitatively, we know that the combination of electric field $E_u$ and magnetic field $B_{ys}$ can lead to the pumping of $X(t)$ through three possible mechanisms: a) direct Hall response via $k_u\rightarrow k_u+E_ut$, b) 2nd order Hall responses mediated by the magnetic field via $k_s\rightarrow k_s+B_{sy}Y(t)$ and $k_y\rightarrow k_y+B_{ys}S(t)$, where $Y(s)$ and $S(t)$ are themselves pumped by $E_u$, and c) 2nd order corrections from phase space renormalization due to the magnetic field.

Rigorously, we have
\begin{eqnarray}
X(t)&=& \int\frac{dX(t)}{dt}dt\notag\\
&=&\frac1{(2\pi)^4}\int_{[0,2\pi]^4} \int[\dot X^u_{\vec k'}(t)+\dot X^s_{\vec k'}(t)+\dot X^y_{\vec k'}(t)]dt\, d^4\vec k'\notag\\
&=&\frac1{(2\pi)^4}\int_{[0,2\pi]^4} [X^u_{\vec k'}(t)+X^s_{\vec k'}(t)+X^y_{\vec k'}(t)]d^4\vec k'\notag\\
&=&\frac1{(2\pi)^4}\int_{[0,2\pi]^4} [(k_u+E_ut)F_{xu}+(k_s+B_{sy}Y^u_{\vec k'}(t))F_{xs}+(k_y+B_{ys}S^u_{\vec k'}(t))F_{xy}]d^4\vec k'\notag\\
&=&\frac1{(2\pi)^4}\int_{[0,2\pi]^4} [(k_u+E_ut)F_{xu}+(k_s+B_{sy}(k_u+E_ut)F_{yu})F_{xs}+(k_y+B_{ys}(k_u+E_ut)F_{su})F_{xy}]d^4\vec k'\notag\\
&=&X(0)+\frac{t}{(2\pi)^4}\int_{[0,2\pi]^4} [E_uF_{xu}+B_{sy}E_uF_{yu}F_{xs}+B_{ys}E_uF_{su}F_{xy}]d^4\vec k'\notag\\
&=&X(0)+\frac{E_ut}{(2\pi)^4}\int_{[0,2\pi]^4} [F_{xu}+B_{sy}F_{yu}F_{xs}+B_{ys}F_{su}F_{xy}]\left[\text{Det}[J]~d^4\vec k''\right]\notag\\
&=&X(0)+\frac{E_ut}{(2\pi)^4}\int_{[0,2\pi]^4} [F_{xu}+B_{sy}F_{yu}F_{xs}+B_{ys}F_{su}F_{xy}]\left[\left(1+B_{ys}F_{ys}\right)d^4\vec k''\right]\notag\\
&\approx &X(0)+\frac{E_ut}{(2\pi)^4}\int_{[0,2\pi]^4}F_{xu}d^4\vec k''+\frac{B_{ys}E_ut}{(2\pi)^4}\int_{[0,2\pi]^4} [F_{xu}F_{ys}-F_{xs}F_{yu}+F_{xy}F_{su}]d^4\vec k''\notag\\
&=&X(0)+\frac{C^{xu}_1}{2\pi}E_ut+\frac{C_2}{(2\pi)^2}B_{ys}E_ut
\end{eqnarray}
The above derivation started from considering all possible momentum-resolved wavepacket contributions to $X(t)$, and arrived at the final results via repeated applications of Eq.~\ref{2D}. $\text{Det}[J]= \sqrt{1+B_{\mu\nu}F_{\mu\nu}}\approx 1+B_{ys}F_{ys}$ is the determinant of the Jacobian $J$ of the transformation between physical and canonical phase space, which are different in the presence of magnetic fields~\cite{lee2018electromagnetic,petrides20186d}. Note that higher order corrections must disappear in 4D systems due to antisymmetry.

\subsubsection{4D Wannier center response}

In generic (entangled) 4D case, the averaged wavepacket center $X(t)$ \emph{does not} describe Wannier center flow, whose nonlinear Hall coefficient is $\frac1{(2\pi)^8}\int_{[0,2\pi]^8}[F_{xu}(\vec k')F_{ys}(\vec k'')-F_{xs}(\vec k')F_{yu}(\vec k'')+F_{xy}(\vec k')F_{su}(\vec k'')]d^4\vec k'd^4\vec k''$. This can be shown by repeating the steps above, but no longer assuming that each Berry curvature depends only on two out of the four momentum coordinates. With each entangled Berry curvature depending on all four momentum indices, we arrive at an 8-dimensional integral.

This is in contrast to the case of unentangled (factorizable) 4D states that can be expressed as a tensor product of 2D QH states in $x$-$s$ and $y$-$u$ space, where only products like $F_{xs}F_{yu}$ exist in the integrand of $C_2$, i.e. $\frac{C_2}{(2\pi)^2}=\frac1{(2\pi)^4}\int_{[0,2\pi]^4}[F_{xu}(\vec k')F_{ys}(\vec k')-F_{xs}(\vec k')F_{yu}(\vec k')+F_{xy}(\vec k')F_{su}(\vec k')]d^4\vec k'd^4\vec k''$.

In such an unentangled case, the Wannier center flow subject to an electric field $E_u$ and magnetic field $B_0=B_{sy}$ is
\begin{equation}
X_{untangled}^{B_0}(t)=P_X(k_u(t))+P_X(B_{sy}P_Y(k_u(t)))
\label{pumping}
\end{equation}
where $k_u=E_ut$ by minimal coupling, and $P_X$, $P_Y$ are the expressions for Wannier polarization in the $X$ and $Y$ directions. To isolate the 2nd Chern number term from Eq. \ref{pumping}, one can take differences to obtain
\begin{equation}
X_{untangled}^{B_0}(t)-X_{untangled}^{-B_0}(t)=2P_X(B_{sy}P_Y(k_u(t)))
\label{pumping2}
\end{equation}

\section{G. Stability Analysis of 4D QH states\label{AppG}}

In this part we will analyze the stability of our system. The deviation to the theoretical values in the experiment is unavoidable, which could happen to the size of the rods, the place where the rods are situated and the angle that the rods rotate to. Hence it's necessary to analyse the stability of our system which robustness of the topological phases. In the meantime, disorder is an indispensable ingredient for some topological phases~\cite{Qin2016Disorder,PhysRevB.96.205304} and our system provides a way to intentionally introduce "designed" disorder.
\begin{figure}[H]
\centering
\includegraphics[width=.8\linewidth]{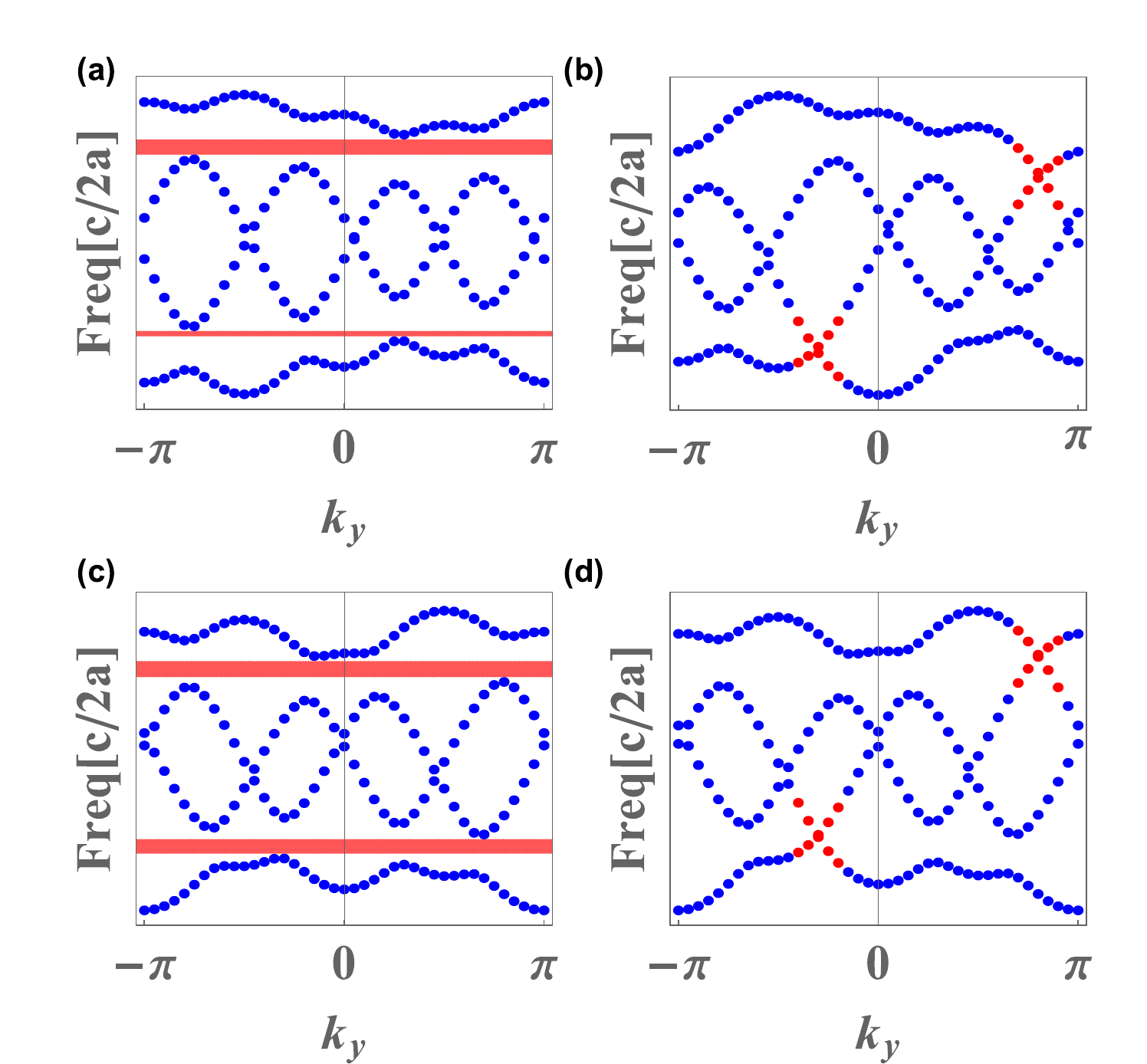}
\caption{The simulation results by COMSOL Multiphysics of the energy band when (a) $\sigma=0.01a$ and a periodic boundary condition is taken. As we can see the bottom bulk band gaps merely exist. (b) $\sigma=0.01a$ and an open boundary condition is taken. Apparently part of the edges state have sunk below the bulk gap. (c) $\sigma=0.005a$ and a periodic boundary condition is taken, and the bulk band gaps still exist. (d) $\sigma=0.005a$ and an open boundary condition is taken. The edge states are located in the bulk band gaps.}
\label{SA}
\end{figure}

We will take the system in Fig.~\ref{fig:2a_new} as an example. As we know, the topological effect in the system is caused by the well designed size of the defects. Thus it's natural and reasonable to focus on the deviations of the defect rods. In order to simulate the deviations, we add a randomness term to every radius of the defect rods as Eq.\ref{stability} shows, all of which obey a Gaussian distribution. Then the problem turns out to become finding the critical value of the standard deviation (SD) of this Gaussian distribution which reveals us how stable the system can be, given the mean value of the Gaussian distribution zero. We adopt several groups of the randomness terms and observe the statistical results to find the critical value.

\begin{align}
r(x) & = r_0 + r_1 cos(2 \pi b x + k_y) + r_d(x) \nonumber \\
r_d(x) & \sim N(\mu,\sigma^2), \mu=0
\label{stability}
\end{align}

The simulation result in Fig.\ref{SA} given by COMSOL Multiphysics shows the critical value of $\sigma$ is around $0.005a$ with $r_0=0.1a$ and $r_1=0.03a$, in which case the bulk gap exists and the edge states doesn't disappear with every group of randomness terms. When $\sigma$ increases to $0.01a$, the bulk gap disappears in half of the groups and they vanish in every group completely when $\sigma$ reaches $0.02a$. It's worth noticing that when the bulk gap disappears, the Chern number can still be defined via an adiabatic process. Thus we can still measure it through topological pumping.

\section{H. 4D photonic nodal lines in a nonsymmorphic lattice\label{AppH}}
\subsection{Tight-binding model\label{AppH1}}
\begin{figure}
    \centering
   \includegraphics[width=0.99\textwidth]{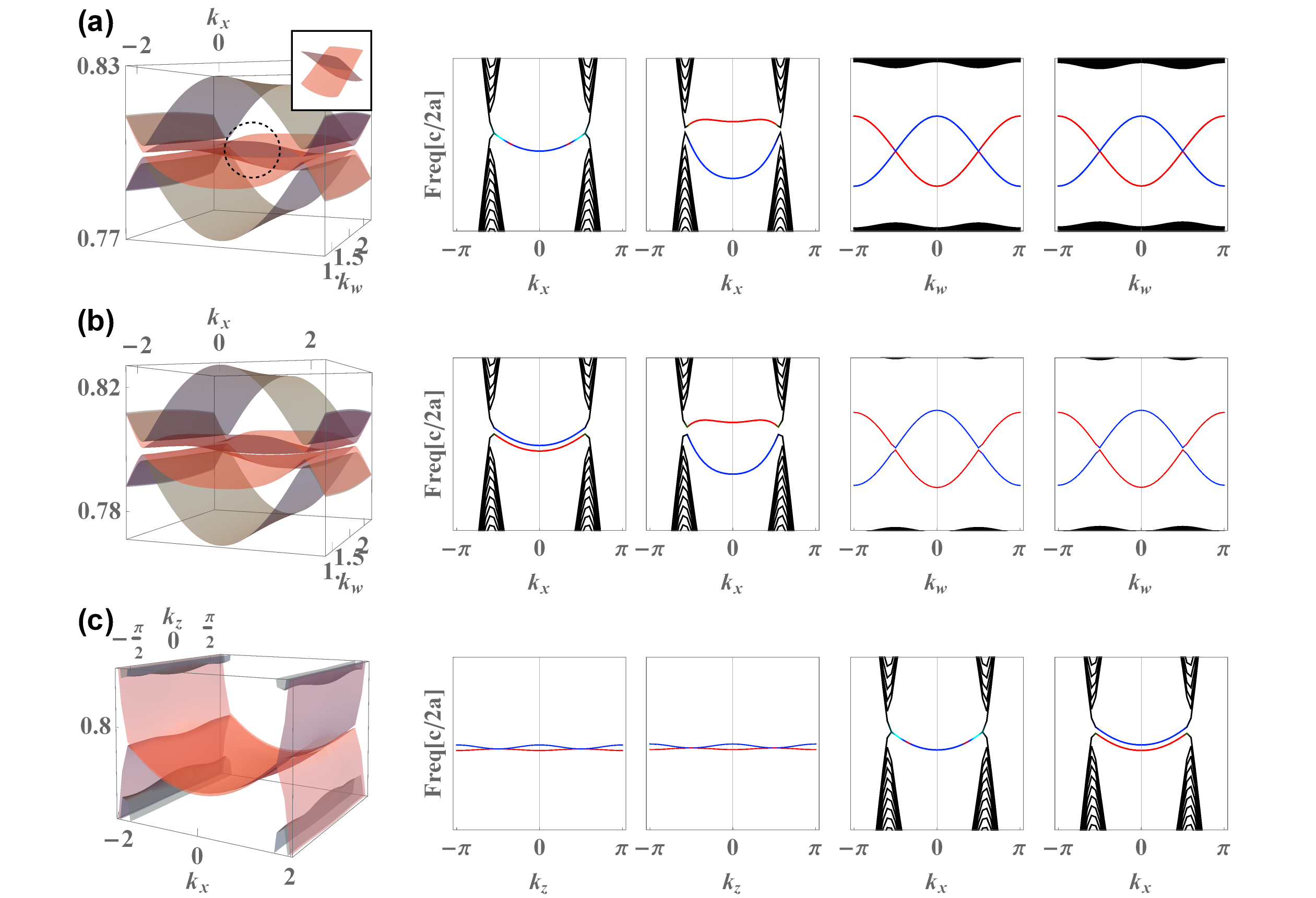}
    \caption{ Edge state on different Brillioun zone crossections from tight-binding model. The open boundary condition is taken on $k_y$ direction, and the z-axis is the frequency $\omega$ ($c/2a$). The two bulk bands constituting the nodal lines are shown in gray color and the boundary modes are sanwitched in between. (a) Left: Open boundary states on $k_x-k_w$ plane where $k_z$ is fixed to be $\pi/2$. Right: The four sub-plots shows the $k_w=\pi/2$; $k_w=\pi/2+0.3$; $k_x=0$; $k_x=0.1$ slices of the energy bands repectively.(b) Left; Open boundary states on $k_x-k_w$ plane where $k_z$ is fixed to be $\pi$. Right: The four sub-plots shows the  $k_w=\pi/2$; $k_w=\pi/2+0.3$; $k_x=0$; $k_x=0.1$ slices of the energy bands repectively. (c) Left; Open boundary state on $kx-kz$ plane where $k_w$ is fixed to be $\pi/2$. Right: The four sub-plots shows the $k_x=0$; $k_x=0.1$; $k_z=\pi/2$; $k_z=\pi$ slices of the energy bands repectively. }
\label{OBC}
\end{figure}

Here we illustrate the tight-binding model for 4D nonsymmorphic pg lattice in the main text. We denote the Hamiltonian as $H(k_x,k_y,k_z,k_w)$ and the glide reflection operation matrix as $U(k_x,k_y)$. For the Hamiltonian to possess the glide reflection symmetry, $H$ should satisfy $UH(k_x,k_y,k_z,k_w)U^\dagger=H(k_x,-k_y,k_z,k_w)$. Under the basis of $|p_x,i\rangle,|p_y,i\rangle$ where $i$ is taken from A to G as shown in Fig.2(a) in the main text, the explicit form of $H$ is as follows,
\begin{equation}
\begin{aligned}
H&=H_d+H_s+H_{onsite}+E_0 \mathbb{I}\\
H_{onsite}&=\lambda \cos{(\pi x+k_z)}+\lambda \cos{(\pi y+k_w)}\\
H_d&=\left(
\begin{array}{c|c}
h_d(-t_1,t_2) & h_a\\
\hline
h_a^* & h_d(-t_3,t_4)
\end{array}
\right)\\
h_d(c_1,c_2)&=\left(
\begin{array}{cccccccc}
0 & c_1 & 0 & c_1 e^{-2ik_xa} &0 & c_2 & 0 & c_2 e^{-2ik_xa}\\
c_1 & 0 &c_1 &0 & c_2 e^{2ik_ya} &0 &c_2 e^{2ik_ya} &0\\
0 & c_1 & 0 & c_1 & 0 & c_2 & 0 & c_2\\
c_1 e^{2ik_xa} &0 & c_1 &0 &c_2 e^{i2(k_x+k_y)a} &0 & c_2 e^{i2k_ya} & 0\\
0 & c_2 e^{-2ik_ya} &0 &c_2 e^{-i2(k_x+k_y)a}& 0 & c_1 & 0 & c_1 e^{-2ik_xa}\\
c_2 & 0 & c_2 &0 &c_1 &0& c_1 &0\\
0 & c_2 e^{-2ik_ya} &0 & c_2 e^{-2ik_ya} &0 & c_1 & 0 & c_1\\
c_2 e^{2ik_x a} & 0 & c_2 &0 & c_1 e^{i2k_xa} &0 & c_1 &0
\end{array}
\right)
\\
h_a&=\left(
\begin{array}{cccccccc}
0 & -r_1 & 0 & r_2 e^{-2ik_xa} &0 & r_4 & 0 & -r_3 e^{-2ik_xa}\\
-r_2 & 0 &r_1 &0 & r_3 e^{2ik_ya} &0 &-r_4 e^{2ik_ya} &0\\
0 & r_2 & 0 & -r_1 & 0 & -r_3 & 0 & r_4\\
r_1 e^{2ik_xa} &0 & -r_2 &0 &-r_4 e^{i2(k_x+k_y)a} &0 & r_3 e^{i2k_ya} & 0\\
0 & r_4 e^{-2ik_ya} &0 &-r_3 e^{-i2(k_x+k_y)a}& 0 & -r_1 & 0 & r_2 e^{-2ik_xa}\\
r_3 & 0 & -r_4 &0 &-r_2 &0& r_1 &0\\
0 & -r_3 e^{-2ik_ya} &0 & r_4 e^{-2ik_ya} &0 & r_2 & 0 & -r_1\\
-r_4 e^{2ik_x a} & 0 & r_3 &0 & r_1 e^{i2k_xa} &0 & -r_2 &0
\end{array}
\right)
\end{aligned}
\end{equation}

\begin{equation}
\begin{aligned}
&H_s=\left(
\begin{array}{c|c}
h_s(s_1,s_2,e_0) & 0\\
\hline
0 & h_s(s_5,s_6,-e_0)
\end{array}
\right)\\
&s_x=1+e^{2ik_xa},s_y=1+e^{2ik_y a}\\
&h_s(c_1,c_2,e_0)=
\left(
\begin{array}{cccccccc}
-e_0/2 & 0& c_1 s_x^* & 0& c_2s_y &0&0 &0\\
0 &-e_0/2 &0 & c_1s_x^* &0 & c_2 s_y  &0&0\\
c_1 s_x &0&-e_0/2&0&0 &0 &c_2 s_y  &0\\
0 & c_1 s_x& 0&-e_0/2&0&0&0& c_2 s_y \\
c_2 s_y^* & 0& 0&0&-e_0/2&0 & c_1 s_x^* &0\\
0 & c_2 s_y^*&0&0 &0 &-e_0/2 &0 & c_1 s_x\\
0 & 0 & c_2 s_y^* &0& c_1 s_x &0 &-e_0/2 &0\\
0 & 0 & 0 & c_2 s_y^* & 0 & c_1 s_x^* &0 &-e_0/2
\end{array}
\right)
\end{aligned}
\end{equation}
where $\lambda=0.012,e_0=0.135,E_0=0.83,s_1=s_2=0.0075,s_5=-0.00125,s_6=0.01,t_1=0.0425,t_2=0.0225,t_3=-0.0825,t_4=-0.0025,r_1=0.0175,r_2=0.04,r_3=0.005,r_4=0.035$, by fitting the tight-binding model with results from the COMSOL Multiphysics simulation.

Under the same basis, U matrix is,

\begin{equation}
\begin{aligned}
&f_x=e^{-2ik_xa},f_y=e^{-2ik_ya}\\
&U(2,1)=U(3,2)=U(4,3)=f_y,U(10,9)=U(11,10)=U(12,11)=-f_y\\
&U(1,4)=-U(9,12)=f_x f_y,U(5,8)=-U(13,16)=f_x\\
&U(6,5)=U(7,6)=U(8,7)=1,U(14,13)=U(15,14)=U(16,15)=-1
\end{aligned}
\end{equation}

With the explicit form of $H$ and $U$, we easily find that $UH(k_x,k_y,k_z,k_w)U^\dagger=H(k_x,-k_y,k_z,k_w)$ holds when $k_z=\pm\pi/2$. Illustrated in Fig.\ref{OBC} are the edge states of 4D pg lattice in different momentum planes where the open boundary is in $k_y$ direction.
\subsection{FEM simulation set-up\label{AppH2}}
The parameters used for COMSOL simulation are as follows. In each subcell, the center of a photonic crystal rod is displaced from the center of the subcell by $(a/4,a/6)$. The major axis for the oval rod is $r_l=1.8r(r_0+r_1\cos{(\pi x+k_z)}+r_2\cos{(\pi y+k_w)})$ and the minor axis is $r_s=r_l/3$, where we take $r=a,r_0=0.1,r_1=r_2=0.035$. The rods A,C,E,G are rotated by -12 degree and B,D,F,H by 12 degree. The relative dielectric constant are all set to be 12. To break the $pg$ symmetry, we set the rotation angle of A,C,E,F to be 30 degree. To simulate the edge states, precisely regulated round rods are placed at the boundary to produce an interface, the radius of which are $0.2a$ with relative dielectric constant being 64. Meanwhile the displacement of the round rod center to the center of subcell is the same as oval rods.

\end{widetext}

\bibliography{apssamp}

\end{document}